\documentclass{svjour3}                     
\smartqed  
\usepackage{graphicx}
\begin{document}

\title{Instruments of RT-2 Experiment onboard CORONAS-PHOTON and their test and evaluation II: 
RT-2/CZT payload
\thanks{This work was made possible in part from a grant from Indian Space Research Organization
(ISRO). The whole-hearted support from G. Madhavan Nair, Ex-Chairman, ISRO, who initiated the RT-2 
project, is gratefully acknowledged.}
}

\titlerunning{RT-2/CZT: Cadmium Zinc Telluride (CZT) detector \ldots..}

\author{Tilak B. Kotoch \and Anuj Nandi \and D. Debnath \and J. P. Malkar \and A. R. Rao \and 
M. K. Hingar \and Vaibhav. P. Madhav \and S. Sreekumar \and Sandip K. Chakrabarti
}

\authorrunning{Kotoch et al.}

\institute{T. B. Kotoch, A. Nandi$^+$, D. Debnath \at
Indian Centre for Space Physics, 43 Chalantika, Garia Station Rd., Kolkata 700 084, India\\
Tel.: +91-33-24366003\\
Fax: +91-33-24622153 Ext. 28\\
         \email{ tilak@csp.res.in; anuj@csp.res.in; dipak@csp.res.in}
  ($+$: Posted at ICSP by Space Science Division, ISRO Head Quarters, Bangalore, India)
         \and
          J. P. Malkar, A. R. Rao, M. K. Hingar, V. P. Madhav, \at
Tata Institute of Fundamental Research, Homi Bhaba Road, Colaba, Mumbai 400 005, India \\
        \and
	      S. Sreekumar \at
	  Vikram Sarabhai Space Centre, VRC, Thiruvanthapuram 695 022, India\\
        \and
              S. K. Chakrabarti \at
              S.N. Bose National Centre for Basic Sciences, JD Block, Salt Lake, Kolkata 700 097, 
India \\
 (Also at Indian Centre for Space Physics, 43 Chalantika, Garia Station Rd., Kolkata 700 084.)\\
	                    Tel.: +91-33-23355706\\
	                  Fax: +91-33-23353477\\
	                \email{chakraba@bose.res.in}           
}

\date{Received: date / Accepted: date}

\maketitle

\begin{abstract}

Cadmium Zinc Telluride (CZT) detectors are high sensitivity and high resolution devices 
for hard X-ray imaging and spectroscopic studies. The new series of CZT detector modules 
(OMS40G256) manufactured by Orbotech Medical Solutions (OMS), Israel, are used in the 
RT-2/CZT payload onboard the CORONAS-PHOTON satellite. The CZT detectors, sensitive in the
energy range of 20 keV to 150 keV, are used to image solar flares in hard X-rays. Since these modules are 
essentially manufactured for commercial applications, we have carried out a series of 
comprehensive tests on these modules so that they can be confidently used in space-borne 
systems. These tests lead us to select the best three pieces of the `Gold' modules 
for the RT-2/CZT payload. This paper presents the characterization of CZT modules and the criteria followed 
for selecting the ones for the RT-2/CZT payload. 

The RT-2/CZT payload carries, along with three CZT modules, a high spatial resolution CMOS 
detector for high resolution imaging of transient X-ray events. Therefore, 
we discuss the characterization of the CMOS detector as well.


\end{abstract}

\section{Introduction}

The RT-2 Experiment onboard the CORONAS-PHOTON satellite (Kotov et al. 2008, Nandi et al. 
2009) is a dedicated experiment for hard X-ray study of solar flares. It consists of three 
main detector payloads, namely RT-2/S, RT-2/G (both NaI(Tl)/CsI(Na) scintillator-Phoswich 
detectors) and RT-2/CZT (solid-state imaging detector) along with one processing electronics 
device, RT-2/E. Detailed descriptions of the Phoswich detectors and processing electronic 
device are given in Debnath et al. (2010) and Sreekumar et al. (2010) and the background 
simulations of the detectors using GEANT-4 toolkit are presented in Sarkar et al. (2010).

The RT-2/CZT payload consists of two different types of imaging detectors: three CZT 
(Cadmium Zinc Telluride) detectors and one CMOS (Complementary Metal Oxide Semiconductor) 
detector, arranged in a configuration of 2 x 2 array. The entire detector assembly (CZT 
and CMOS) sits below a collimator ($\sim$32 cm height) with two different types of masking 
devices, namely Coded Aperture Mask (CAM) and Fresnel Zone Plate (FZP). The payload 
configuration, simulation and experimental results with the coding devices are discussed in 
Nandi et al. (2010). Due to the different dimensions of the implemented coding devices, this 
payload has a Field of View (FOV) from $\sim$ 6$'$ to 6$^\circ$. RT-2/CZT payload is the 
only imaging device onboard the CORONAS-PHOTON satellite to image solar flares in hard 
X-rays between $20$ keV to $150$ keV. The CZT detectors have good spectral resolution but
moderate spatial resolution. The CMOS detector is capable of imaging 
with a high resolution. However, it is a single channel device,
and is not capable of generating a spectrum.

RT-2/CZT payload is placed outside the hermetically sealed vessel of the satellite and 
co-aligned to the Sun pointing axis. In \S 2 and \S 3 we present individual detector module 
testing and selection criteria of CZT and CMOS detectors for the flight use. In \S 4 we 
present the test set-up for the payload testing and results at different conditions and in 
\S 5, we make concluding remarks.

\section {CZT detector testing and selection criteria}

Cadmium Zinc Telluride (CZT) is an extrinsic semiconductor X-ray detector with a band gap 
energy of 1.5 - 2.0 eV and an average atomic number, $Z\sim50$. Due to the high band gap 
energy, it can be operated at room temperature. As the atomic number ($Z$) is high, even a few mm of CZT 
can absorb hard X-rays efficiently (Knoll 1999). In comparison to 
other X-ray detectors like scintillation detector and proportional counter, CZT needs a 
relatively low energy deposition for electron-hole pair formation resulting in a good energy 
resolution (Kotoch et al. 2008). CZT detectors have other major benefits, for example, it can 
be pixilated by further dividing the sensitive area with segmented anode into multiple pixels 
which leads to spectral enhancement due to small pixel effect. The current state of art allows 
one to develop large area X-ray detectors in the form of a mosaic of CZT pixel devices, making 
this technology really appealing for use in hard X-ray astronomy.

CZT detectors already have been successfully used in the Swift satellite (Barthelmy et al. 
2005) and in recently launched Indian Moon mission Chandrayaan-1 (Vadawale et al. 2009). The 
same detectors would be used in the forthcoming missions like ASTROSAT, EXIST and 
Constellation-X. We have used CZT detectors in the RT-2 Experiment for hard X-ray solar flare 
studies. Three CZT detector modules are placed in RT-2/CZT, one of the main payloads of the 
RT-2 Experiment onboard CORONAS-PHOTON mission, which was launched successfully from Russia on 
January 30, 2009.

The RT-2 CZT detector modules (OMS40G256) were procured from Orbotech Medical Solutions Ltd., 
Israel. The detector module (Fig. 1) consists of a 5 mm thick CZT crystal having dimension of 
$3.9 \times 3.9$ cm$^2$. Each module is pixilated to have $16 \times 16$ pixels, where each 
inner pixel size is $2.46 \times 2.46$ mm$^2$, while the edge pixels have a pitch of $2.28 
\times 2.28$ mm$^2$. The CZT crystal composition (i.e., Cd$_{0.9}$Zn$_{0.1}$Te) and growth are 
achieved using the MHB (Modified horizontal Bridgman) technique (Yadav et al. 2005, Jung et al. 
2007, Vadawale et al. 2009). It has a density of $5.85$ gm/cm$^3$. The crystal is N-type 
conductive having electron as a major charge carrier. The electrodes contacts are made of 
Indium (Lachish et al. 1999). Cathode is a single mono-electrode, whereas the anode is 
pixilated to pad size of $1.86 \times 1.86$ mm$^2$ with 0.42 mm gap. The CZT detector crystal has been 
integrated with a multi-channel read-out ASIC's having 128 channel which are self-triggered and 
data driven (Yadav et al. 2005, Jung et al. 2007, Vadawale et al. 2009). Its threshold can be 
externally controlled. These modules can be easily mounted and dismounted from the motherboard 
through two 20-pin surface mounted connectors in the back side.

OMS40G256 detectors are capable of detecting X-rays of energies ranging from $10 - 200$ keV. 
In addition to having all digital interfaces, an optional temperature measurement sensor is 
integrated with the front end electronics to monitor the temperature of the module, which is 
useful in monitoring the heat dissipation by the ASICs. The ASIC in each CZT module contains its own 
pixel map information and inbuilt event storing memory (FIFO) for up to 256 events. The 
peak position shift of any source spectrum is less than 0.1 keV/$^{\circ}$C. The overall power 
consumption of a module is about 300 mW, which is about $\sim 50$ \% lower than that of the 
earlier version of the series of products (Yadav et al. 2005, Jung et al. 2007, Vadawale et al. 2009). 

A detailed study of these detectors was carried out for finding its performance, efficiency, 
etc. in various environmental conditions and to authenticate its use for the space environment. 

\begin{figure}
\vbox{
\vskip 0.0cm
\centerline{
   \includegraphics[width=3.5cm, height=3.5cm]{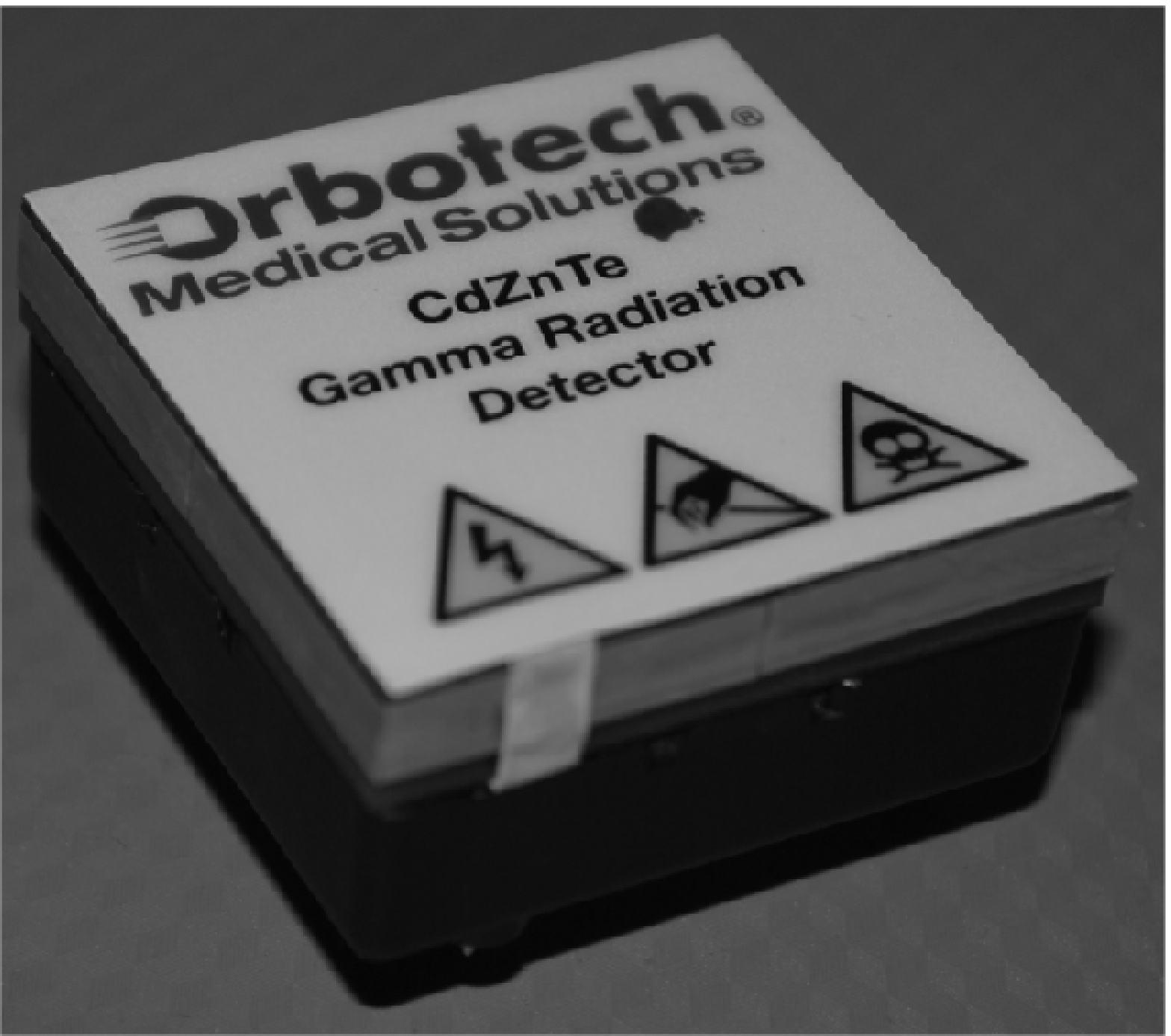}
   \includegraphics[width=3.5cm, height=3.5cm]{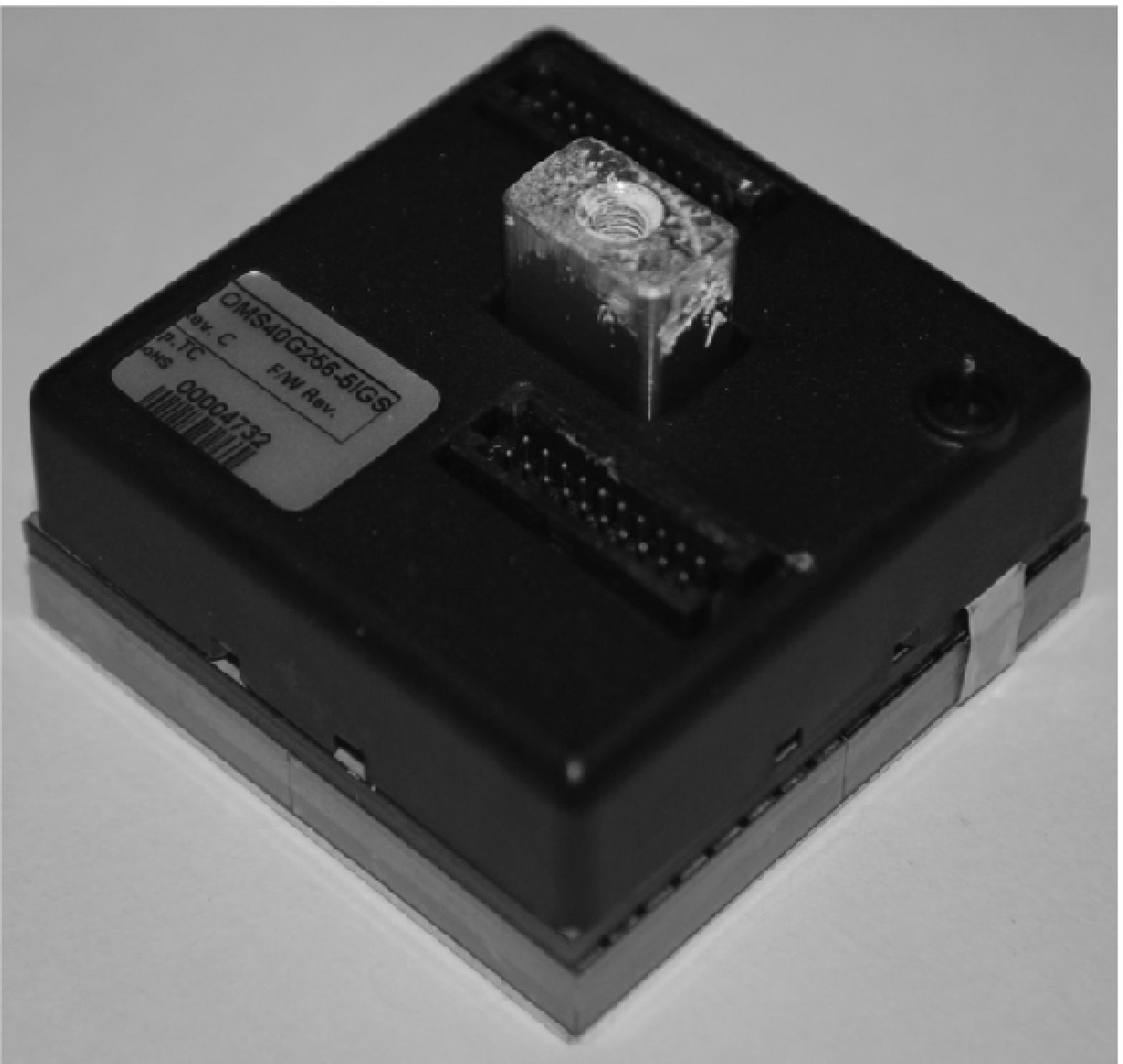}}
\vspace{0.0cm}
\centering \caption{A CZT module of dimension 3.96 cm $\times$ 3.96 cm $\times$ 5 mm 
(OMS40G256) (a) Left panel shows the isometric view and (b) right panel shows the bottom view.
}}
\label{}
\end{figure}

\subsection {Experimental Setup}

The System Development Kit (SDK), developed by Orbotech, was used to test our sample detector 
modules. This system is available from Orbotech commercially and consists of two packages: a 
detector box and a SDK box. The detector box can accommodate up to 20 detector modules (in a 
$4 \times 5$ matrix) and it is connected to the SDK box by a 50 pin cable. It gets HV (High 
Voltage) bias directly from the power supply and the low voltage power through the SDK. The 
SDK is connected to a computer with an accompanied software which basically allows the control 
of all ASIC functionalities and data acquisition from the CZT detector modules (Yadav et al.    
2005; Vadawale et al. 2009). The SDK software tool saves the spectrum of each pixel data in 
ASCII files containing the information of event counts per channel.    

For the tests at low temperatures, we used a regular laboratory cold chamber which can 
actively control temperature between -30$^{\circ}$C to 0$^{\circ}$C with the accuracy of 
1$^{\circ}$C. We have made the detector box air-tight by covering it with a plastic bag 
containing two pouches of dried silica gel. The major source of heat dissipation is at the 
bottom of the CZT detector modules. Therefore, we positioned a temperature probe (thermistor) 
at the cold finger of the detector module in order to control the detector temperature. The 
chamber temperature is maintained and varied with respect to this monitoring thermistor 
temperature. 

\subsection {Test Procedure}

A test procedure was planned on the basis of detector specification provided by OMS Ltd. and 
desired test environment. The schematic block diagram along with the lab setup of the 
experiment is shown in Fig. 2. 

\begin{figure}[h]
\vbox{
\vskip 0.0cm
\centerline{
 \includegraphics[width=7cm, height=5.2cm, angle=0]{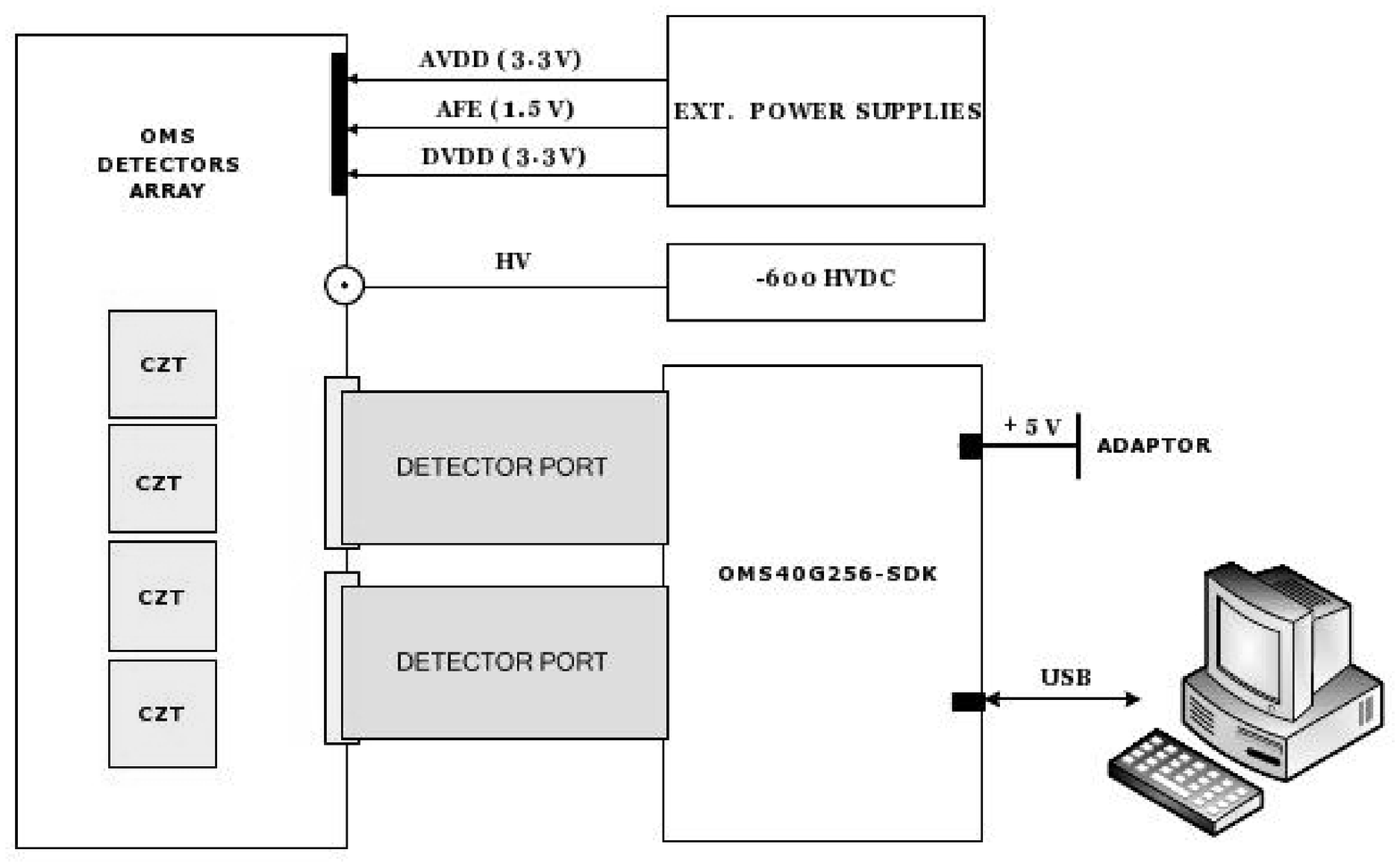}
 \includegraphics[width=7cm, height=5.2cm, angle=0]{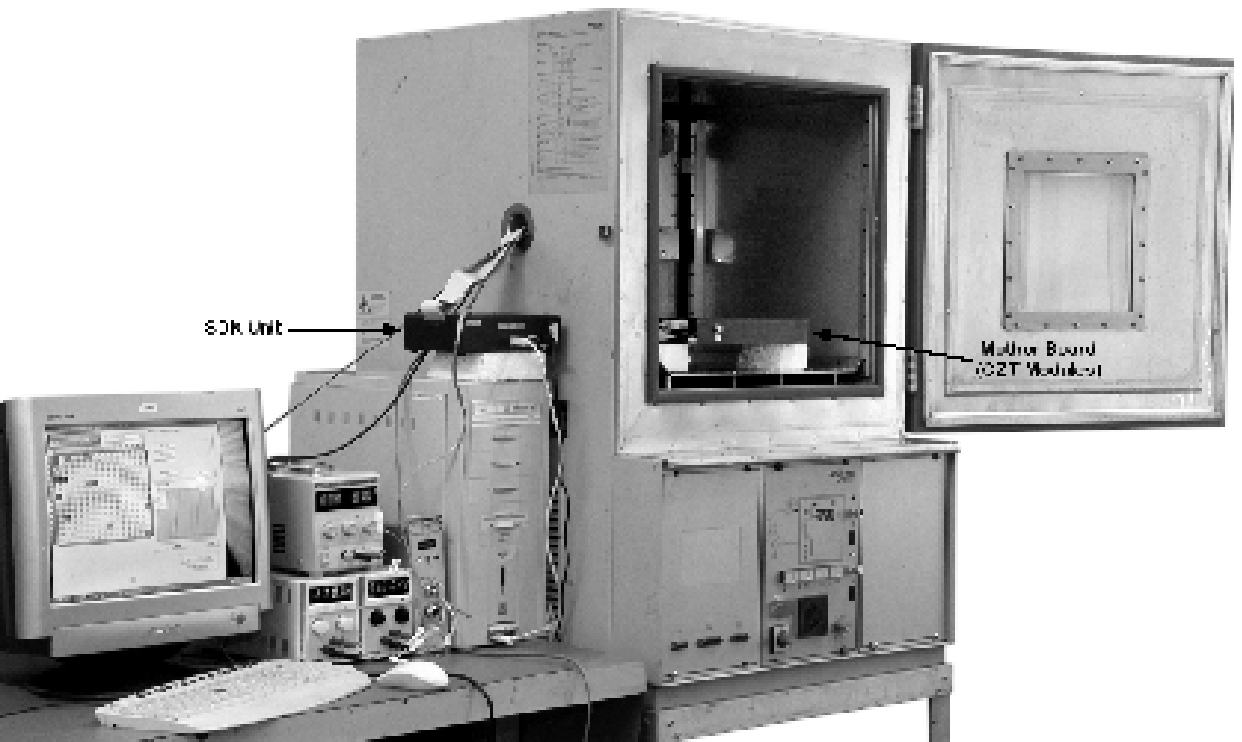}}
\caption{Schematic block diagram of the experimental setup (left) and the lab setup 
(right) for individual CZT module testing.}}
\label{}
\end{figure}

CZT detectors were stored in a low humidity control chamber. This chamber was kept in the 
clean room for mounting or dismounting of CZT modules from the motherboard (detector board). 
Radioactive sources Am$^{241}$ and Cd$^{109}$ were used for various tests and calibration of 
CZT detectors. Three separate power supplies were used for the OMS detector test unit to 
provide two analog (called AVDD and AFE) and one digital (called DVDD) supply voltages along 
with a common grounding. A high voltage power supply (HVPS) with -$600$V is fed at the top of 
CZT detector crystal. During this period, we monitored each power supply to verify the nominal 
power consumption for each detector (Table 1). A dedicated USB port is used for interfacing 
between the system and SDK unit for data acquisition and to control various parameters of the 
detector. 

On accomplishing all the basic settings required for the test, CZT modules were directly 
irradiated with two radioactive sources (Am$^{241}$ and Cd$^{109}$). Appropriate separation 
between the sources and detector was kept in such a way that a uniform illumination occurs 
over the entire detector area. The data were accumulated for intervals of 150 - 1800 seconds 
as required to obtain a good counting statistics (typically several thousand photons under one 
energy peak in the spectrum which gives the statistical accuracy of the peak position 
determination correct to $\sim$1\%). The average threshold of each CZT module was set at 40 
keV in order to avoid low energy noise and to obtain a good spectrum at 59.5 keV and 88.0 keV. 
It was noticed that heating of detector ASIC creates a large number of noisy pixels in the 
detector. To prevent this effect, the detector temperature was maintained below 20$^\circ$C 
inside the cold chamber. In order to avoid the moisture condensation, the entire detector unit 
was kept in airtight container before placing inside the cold chamber. The temperature of 
detector box was maintained with the help of thermistor placed at the base. Tests were 
conducted at various temperatures (-20$^\circ$C to 20$^\circ$C) for characterization (such as, 
overall performance, energy resolution and stability).

\begin{table}[h]
\scriptsize
\centering
\centerline {Table 1: Detector power consumption (for one module)}
\vskip 0.2cm
\begin{tabular}{lccc}
\hline
\hline
Voltage ID & Voltage $\pm$5\% & Typical current & Maximum current\\
\hline
AVDD & +3.3 V & 30 mA & 60 mA\\
AFE  & +1.5 V & 20 mA & 40 mA\\
DVDD & +3.3 V & 40 mA & 80 mA\\
HVPS & -600 V & 20 $\mu$A & 60 $\mu$A\\
\hline
\end{tabular}
\end{table}

\subsection {Module screening criteria}

The CZT modules can be operated in command mode (to set thresholds etc.) or event mode when 
it works as a self-triggered detector and stores pixel number and the energy information (10 
bit ADC) in an in-built FIFO, which can be interrogated and from which the data can be 
collected. The SDK program handles various tasks like communication with CZT digital interface, 
data acquisition and analysis of the data gathered from CZT detectors. It also provides the 
performances of each pixel and the various average parameter values of 256 pixels of the 
selected detector. A single detector (CZT module) can be selected for analysis among the array 
of detectors inside the detector box using the SDK application. It has also the features like 
to set and as well as to enable or disable the threshold of detector parameters, view each 
detector pixel spectrum and events collected by the detector. The information on each pixel can 
be displayed with appropriate color-coding. For example, Fig. 3 shows a sample display with the 
color indicating the pixel quality, quantified by various parameters like the percentage energy 
resolution at FWHM (D[\%]); the efficiency (S[\%]) defined by the ratio between counts under 
the photo-peak ($\pm$ 6.5$\%$ of the peak energy) and the total number of counts under the 
spectrum (integrated from 40 keV to the peak energy + 6.5$\%$).

\begin{figure}[h]
\vbox{
\vskip 0.0cm
\centerline{
   \includegraphics[width=12cm,height=8cm]{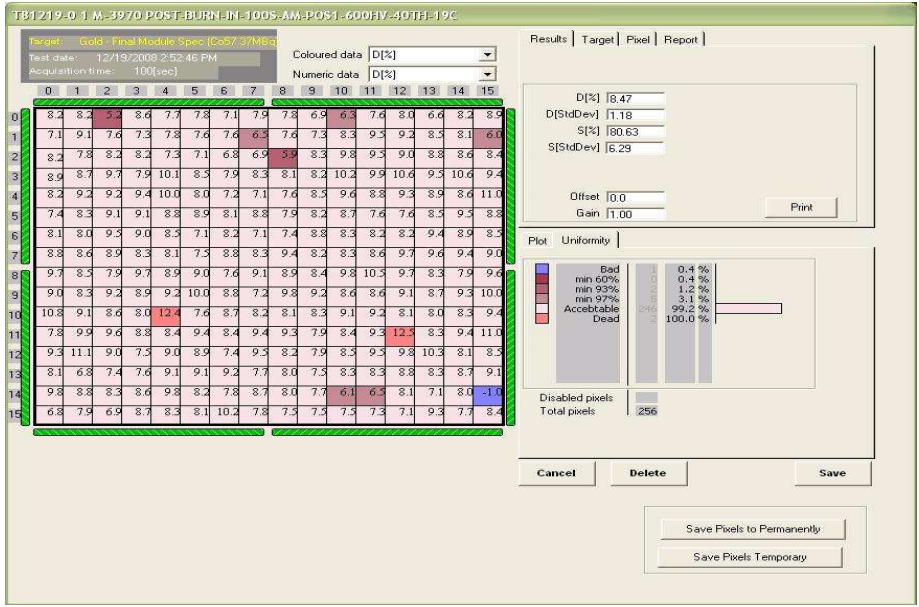}}
\vspace{0.0cm}
\caption{Display shows the color-coded status of pixels of a CZT module from the SDK set-up. 
Status of each pixel is based on various parameters (see text, for details).}}
\label{}
\end{figure}

SDK software application was initially used for data acquisition and to quickly analyze the 
data acquired of the selected module, during the lab test. For a detailed analysis, however, 
we used the SDK software to store data in event mode files to be read by a tool we have 
developed using the Interactive Data Language (IDL). This software was specially designed to 
analyze spectral data with two calibration peaks (e.g. 59.5 keV from Am$^{241}$ and 88 keV 
from Cd$^{109}$). The detector gain (relation between energy and channel) was assumed to be 
linear in the 20 - 200 keV energy range, as given by specifications, and it was found to be 
satisfactory during the CZT modules laboratory tests with the rms deviation from a linear fit 
being $<$0.5\% at 122 keV. Analysis was done by first identifying and removing the unwanted 
noisy, dead and bad pixels (quantitative descriptions of all types of pixels are given in the 
next section). Gaussian fittings were made to the two known photo-peaks. The gain (keV/Channel) 
and offset for individual pixels were calculated using these two photo-peaks. The FWHM derived 
from the Gaussian fitting is called FWHM$_{Fit}$ (fitted). FWHM was also estimated by the 
number of channels between the two edges of the peak at half maximum and this is called 
FWHM$_{Cal}$ (calculated).

During the analysis, we also calculated the peak efficiency and peak counts. The peak 
efficiency is defined as the total counts under peak position within 2$\sigma$ level divided 
by the total counts in the spectrum. We have followed the same procedure for all 30 `Gold' 
(module with atleast 97\% of pixels having average energy resolution of $\sim$6.5\% @ 122 
keV) quality modules that we have selected for initial testing and screening.

\subsection {Room Temperature Test}

Based on the performance (energy resolution, efficiency, stability, etc.) of individual 
modules, we selected 8 CZT modules out of 30 modules for the final screening. At room 
temperature tests, we calculated various parameters (such as peak position, FWHM, Gain, 
Offset, Energy resolution, peak efficiency etc.) for each of the 256 pixels of all 8 CZT 
modules. 

In Fig. 4, we have plotted various parameter of an individual CZT module to demonstrate the 
pixel-to-pixel variation. There are few disabled pixels during data acquisition (dead or noisy 
pixel) and pixels with poor response (bad pixels). These pixels are seen either as a dip or 
high value in comparison to the other pixels in a given detector module. It can be seen from 
the figure that if we exclude these disqualified pixels (which are less than 4 in a given 
module), the overall performance is very good. The periodic dip in the total counts variation 
for all pixels (256) is due to the smaller area of the edge pixels (eg. 16th, 32th, 48th etc.). 
It is noted that two corner pixels show peak values lower than other pixels, which could be 
due to the presence of bad pixels in a module, leading to poor energy resolution. The pixel 
gain (keV/channel) and offset of the entire module appears to remain constant whereas the 
peak channel number (Pulse Height or PH) has maximum variation of 5$\%$. Results shown in Fig. 
4 are from one of the CZT module (No. 2789) that is used in space-flight of RT-2/CZT payload. 

\begin{figure}[h]
\centerline{
\includegraphics[width=12cm, height=8cm]{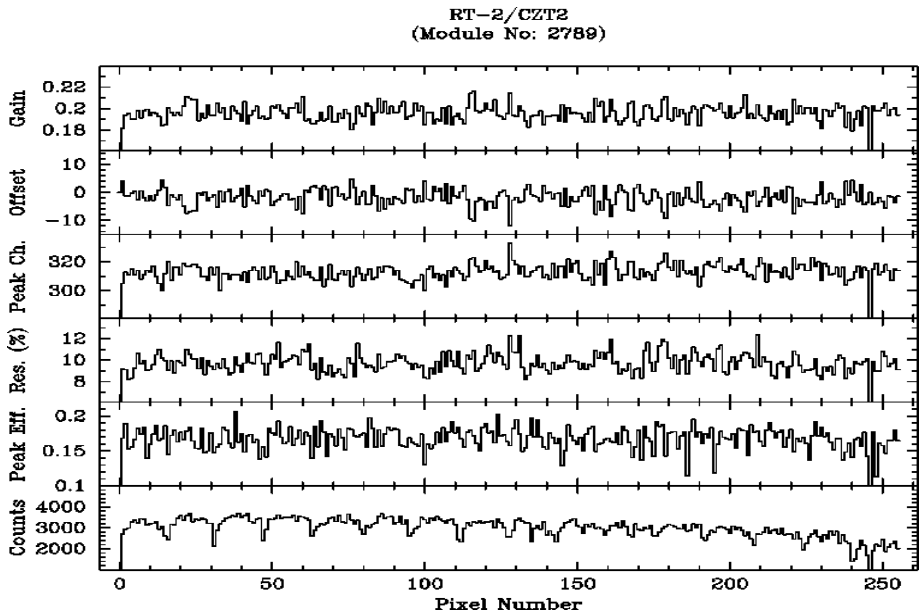}}
\caption{Variation of different parameters of each of the 256 pixels of one CZT module (No. 
2789). See text for details.}
\label{}
\end{figure}

A standard procedure to evaluate any device is to compare the findings with average outcome 
from other similar devices. But in case of CZT modules, it is not appropriate to justify all 
these parameters on the basis of average value calculated by including all pixels of a module, 
as overall outcome will get affected due to presence of disqualified pixels in a module. 
Therefore, a methodology was adopted to resolve this complexity of module evaluation by 
considering the average energy resolution as a prime feature among other parameters and by 
selecting those pixels which have energy resolutions very close to the best energy resolution 
pixel. This was achieved using a code (using IDL) that gives an average energy resolution of 
best 90$\%$ pixels and best 75$\%$ pixels. As a result the disqualified (hot, dead and bad) 
pixels were removed automatically from consideration. In addition, the program also provides 
the information on total number of disqualified pixels along with pixel address. To reduce 
the impact of these disqualified pixels on the average energy resolution of module, they are 
filtered out by the program during the calculation. This leads to a better understanding of 
the module performance and also while evaluating them for the flight purpose. 

Similarly, an average value of all defined parameters (Gain, Offset, FWHM etc.) is calculated 
over best 90$\%$ and 75$\%$ pixels for all 8 CZT detector modules used in laboratory testing 
at room temperature. The variation of the parameters of all 8 CZT modules is shown in Fig. 5. 
The average parameter values of all 8 modules, along with the error bars were calculated and 
they show a great uniformity. The gain and offset of each module is almost consistently around 
0.195 $\pm$ 0.010 keV/channel and -2 $\pm$ 3 channel whereas the energy resolution given in 
the term of FWHM$_{Cal}$ and FWHM$_{Fit}$ is about 9.5 $\pm$ 0.5$\%$ and 10.5 $\pm$ 0.5$\%$. 
In Table 2, we have summarized the energy resolution (at 59.5 keV) informations against pixel 
wise distribution along with the number of disqualified pixels for all the 8 CZT detector 
modules.

\begin{figure}[h]
\centering{\includegraphics[width=8cm, height=12cm, angle=270]{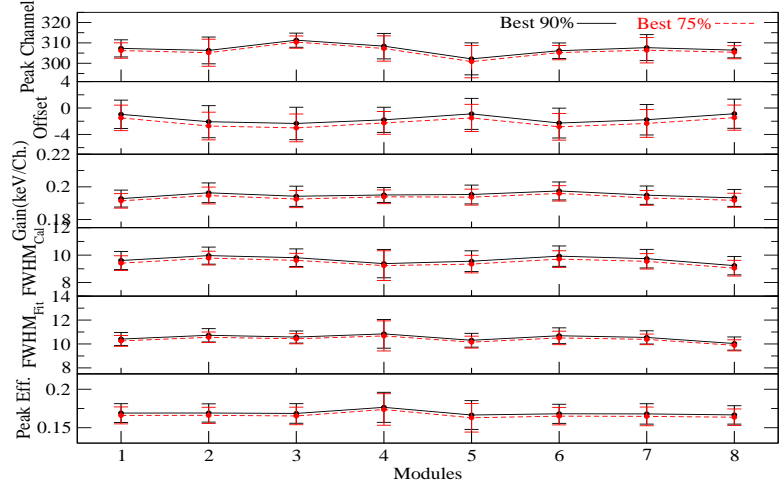}}
\caption{Average value of various parameters along with error bar (standard deviation), for 
the best 75\% and the best 90\% pixels of all 8 CZT modules used in the room 
temperature tests.}
\label{}
\end{figure}

\begin{table}[h]
\scriptsize
\centering
\centerline {Table 2: Pixel wise energy resolution distribution of the CZT modules}
\vskip 0.2cm
\begin{tabular}{|c|c|c|c|c|c|c|}
\hline
\bf{Sr.}&\bf{Module}&\bf{Disqualified }&\multicolumn{2}{c|}{\bf{Pixels with energy}}&\multicolumn{2}{c|}{\bf{Average Energy}}\\
\bf{No.}& \bf{No.}&\bf{Pixels}&\multicolumn{2}{c|}{\bf{resolution @59.5 keV}} & \multicolumn{2}{c|}{\bf{Resolution}} \\
\cline{4-7}
 & & &\bf{~~~$<$10$\%$~~~}&\bf{10-14$\%$}&\bf{Best 90$\%$}&\bf{Best 75$\%$}\\
\hline
1 & 2781 & 2 & 54 & 200 &10.41 $\pm$ 0.54 & 10.26 $\pm$ 0.45\\ 
\hline
2 & 2976 & 2 & 23 & 231 &10.73 $\pm$ 0.56 & 10.56 $\pm$ 0.44\\
\hline
3 & 2789 & 3 & 27 & 226 & 10.58 $\pm$ 0.50 & 10.44 $\pm$ 0.43\\ 
\hline
4 & 2783 & 2 & 15 & 239 & 10.85 $\pm$ 1.20 & 10.67 $\pm$ 1.26\\ 
\hline
5 & 2954 & 5 & 74 & 177 & 10.31 $\pm$ 0.58 & 10.15 $\pm$ 0.50\\ 
\hline
6 & 2957 & 2 & 37 & 217 & 10.69 $\pm$ 0.65 & 10.51 $\pm$ 0.56\\ 
\hline
7 & 2948 & 3 & 41 & 212 & 10.55 $\pm$ 0.55 & 10.38 $\pm$ 0.44\\ 
\hline
8 & 2738 & 5 & 102 & 149 & 10.04 $\pm$ 0.55 & 09.88 $\pm$ 0.47\\
\hline
\end{tabular}
\end{table}

From Table 2, it is clearly seen that the average energy resolution of the CZT detector 
modules lies around 10.5 $\pm$ 0.5$\%$ at 59.5 keV. The performance of modules depends on 
both the number of disqualified pixels and energy resolutions. There are two types of bad 
modules: one is defined by a bad spectroscopic performance (the average energy resolutions), 
while the other type is defined by the presence of large number of disqualified pixels. 
During the test, the number of dead pixels remains constant with the number of readings taken 
at different interval of time whereas the hot and bad pixels show variation in their numbers 
for a given module. There are some hot and bad pixels of the module which can be recovered by 
optimizing grounding of the entire test setup and by proper cooling arrangement for the test 
unit (mother board). 

During the analysis, we have classified the pixel depending on their spectral behaviors in 
the following way:

\begin{itemize}
\item 
\textbf{Hot (Noisy) pixel:} These pixels are the ones having a very sharp peak or 
exponentially descending peak (higher count rate) (Fig. 6) compared to the average peak from 
the source or background. In a module, there are some pixels which are very sensitive to 
noise either from external sources like inappropriate grounding of the test setup or due to 
deficiency in heat transfer process which leads to heating of the CZT modules. 
In Fig. 6, we have shown two spectrum (accumulation time of 150 sec) of different pixels of the
same CZT module. In order to show the enhanced feature of noise effect (hot pixel behavior) 
of two different pixels, we have plotted the spectra in different channel scale (total channel 
numbers 1024). These noisy pixels can be considerably reduced by taking appropriate action 
against the origin of noise. 
\end{itemize}

\begin{figure}
\vbox{
\vskip 0.0cm
\centerline{
   \includegraphics[height=4.8cm, width=5cm]{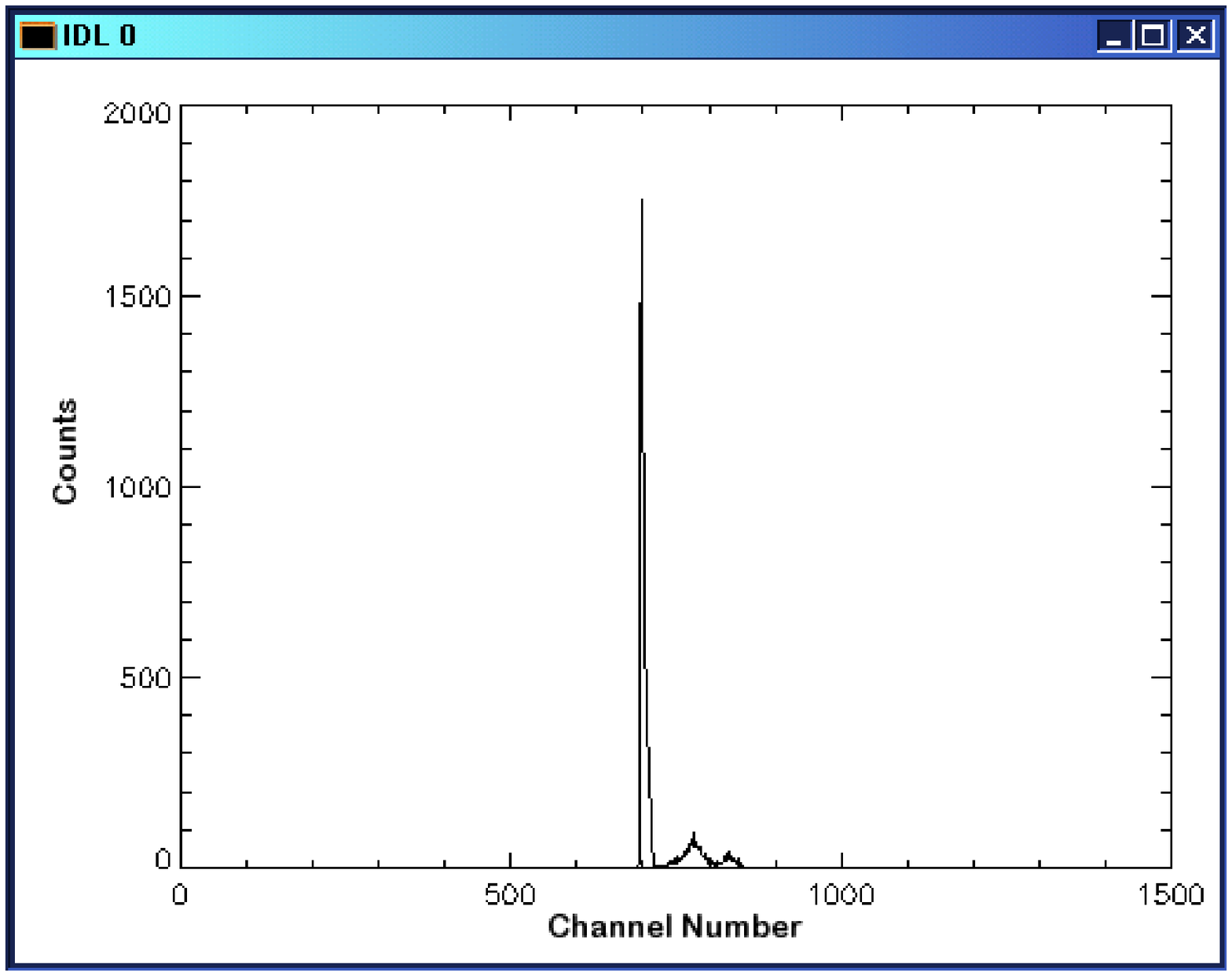}
   \includegraphics[height=4.8cm, width=5cm]{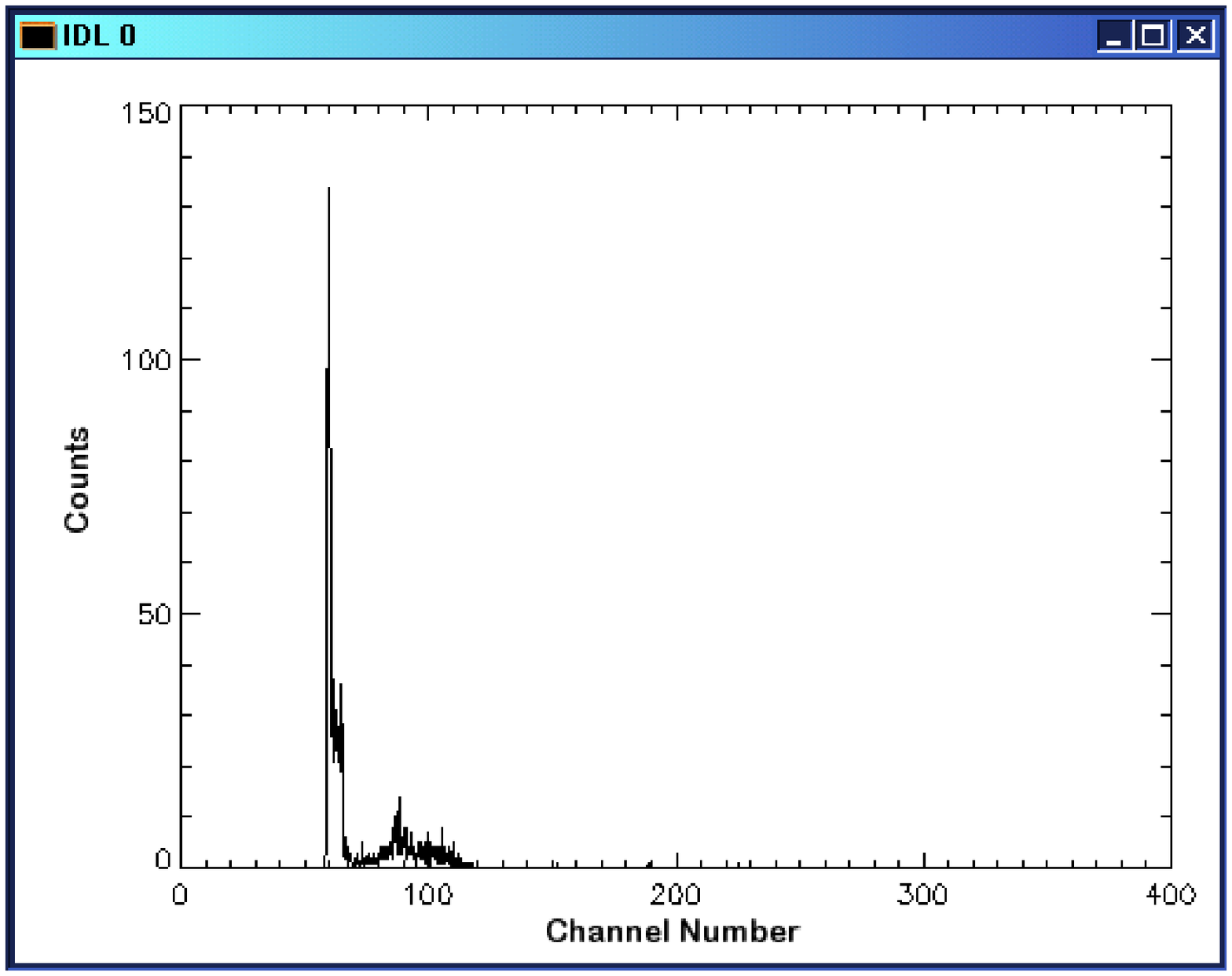}}
\caption{Examples of hot pixel spectra for a Am$^{241}$ source. Data is accumulated for 150 
sec to plot the spectrum in Counts (Y-axis) vs Channels (X-axis).}}
\label{}
\end{figure}

\begin{itemize}
\item 
\textbf{Good pixel:} A pixel having a very good efficiency and a Gaussian shaped photo-peak for 
any radioactive source and uniform background spectrum is defined as a good pixel. Fig. 7 
shows the spectrum containing two peaks obtained from the radioactive sources Am$^{241}$ (59.5 
keV) and Cd$^{109}$ (88.0 keV). The spectrum is obtained for the period of 150 second, similar 
to the hot pixel accumulation time.
\end{itemize}

\begin{figure}[h]
\centerline{
 \includegraphics[height=4.6cm, width=5cm]{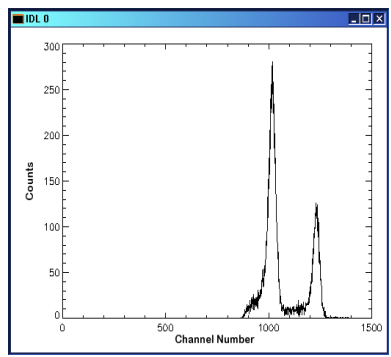}}
\caption{Good pixel: Uncalibrated spectrum from radioactive source Am$^{241}$ (59.5 keV) and 
Cd$^{109}$ (88.0 keV). Spectral data is accumulated for 150 sec.}
\label{}
\end{figure}

\begin{itemize}
\item 
\textbf{Bad pixel:} Pixel having poor sensitivity either due to crystal defect structure 
(during crystal growth) or improper bonding between crystal and the readout devices (ASIC) of 
the module. Also, it might be due to mishandling of the device without having proper ESD 
protection. In comparison to normal (good) pixels, the  average count from the source or 
background are quite less in number ($<$10 counts per 150 sec for photo-peak). An 
illustration of a bad pixel is shown in Fig. 8.
\end{itemize}

\begin{figure}[h]
\vbox{
\centerline{
   \includegraphics[height=4.6cm, width=5cm]{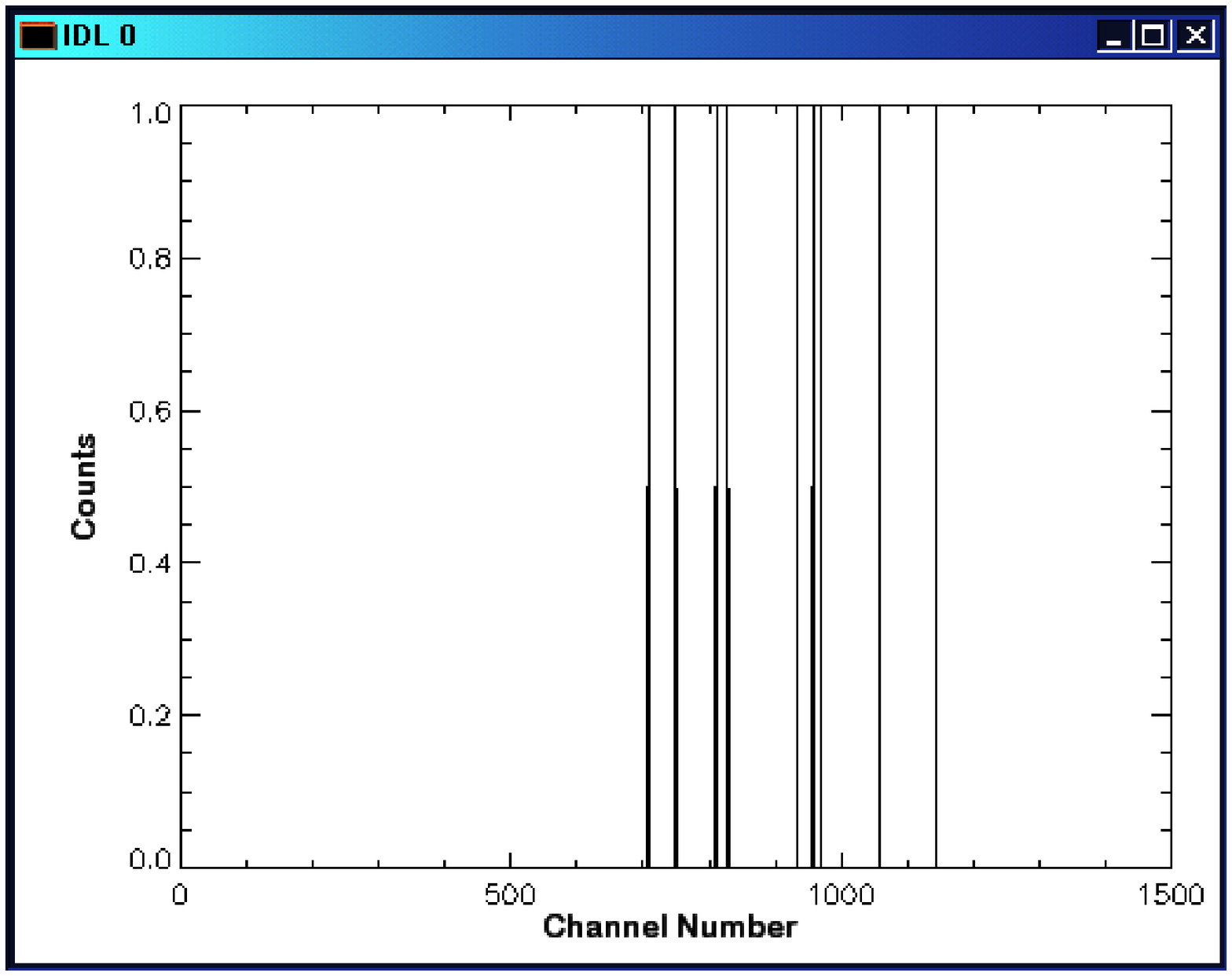}
   \includegraphics[height=4.6cm, width=5cm]{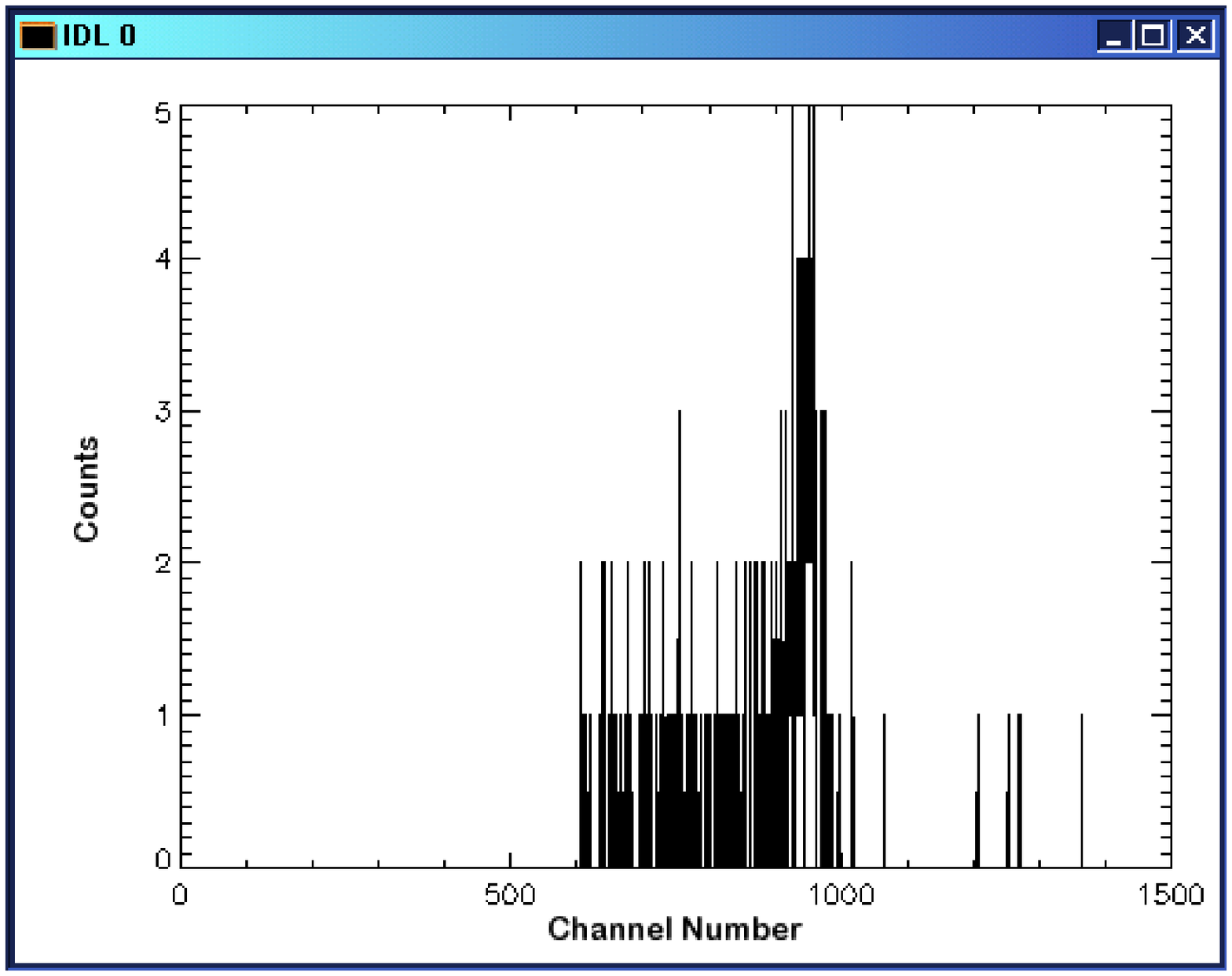}}
\vspace{0.0cm}
\caption{Bad pixels: (a) Left panel shows crystal inherent deficiency effect and (b) Right 
panel shows the ESD negligence effect.}}
\label{}
\end{figure}

\begin{itemize}
\item 
\textbf{Dead pixel:} Pixel with zero output (no count) and having no sensitivity to detect X-rays 
is said to be a dead pixel. There are many possibilities for a pixel being dead, like crystal 
defect, weak bonding between crystal and electrodes, mishandling without ESD protection. 
Unlike the case of a hot (noisy) pixel, 
there is no chance of recovery of a dead pixel by any means.
\end{itemize}

\subsection{Cooling Effect}

At room temperature testing of CZT detector, it was noticed that the modules had an average 
temperature of 5$^{o}$C - 6$^{o}$C higher than the environment temperature, resulting in an 
increase in leakage current which plays a significant role in the response of each detector 
module, especially, on the spectroscopic performance. The key features to monitor are 
performance characteristics of CZT detector and the influence of low temperature operation 
over the leakage current of the semiconductor.

Low temperature (cold) tests were carried out over six CZT modules with the same radioactive 
sources used for previous measurements. A cold test was performed with the operating temperature 
ranging from -20$^\circ$C to +20$^\circ$C and the data was taken when the detector operating 
temperature become stable within $\pm$ 1$^\circ$C. Fig. 9 illustrates the results of cold test 
in terms of average energy resolutions of each module with respect to the operating 
temperature, for the 59.5 keV emission peak from Am$^{241}$ source. These results show clearly 
that the energy resolution of CZT modules improves with decreasing temperature down to 
10$^\circ$C and become saturated at 0$^\circ$C. Below zero degree centigrade, their 
spectroscopic behavior change toward energy resolution similar to the ones measured at room 
temperature. In Fig. 19, we have also shown the result from a flight CZT module, which also shows
the same spectroscopic behaviour.

\begin{figure}[h]
\centerline {
 \includegraphics[width=12cm, height=8cm]{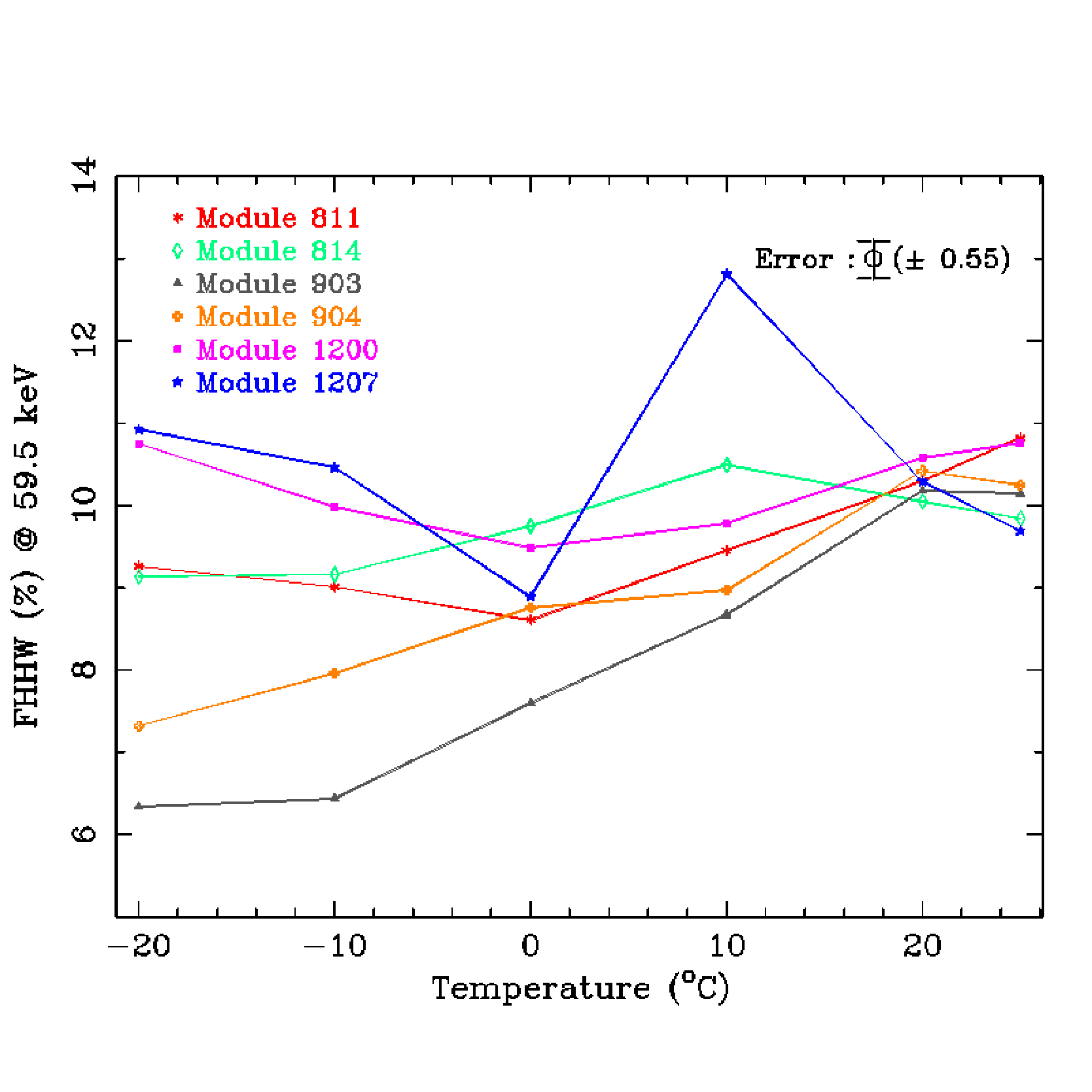}}
\vskip 0.1cm
\caption {Variation of the average energy resolution (59.5 keV emission peak from Am$^{241}$) 
with operating temperature for six CZT modules. Typical error bar (maxi.) is marked on the figure.}
\label{}
\end{figure}

\begin{figure}[h]
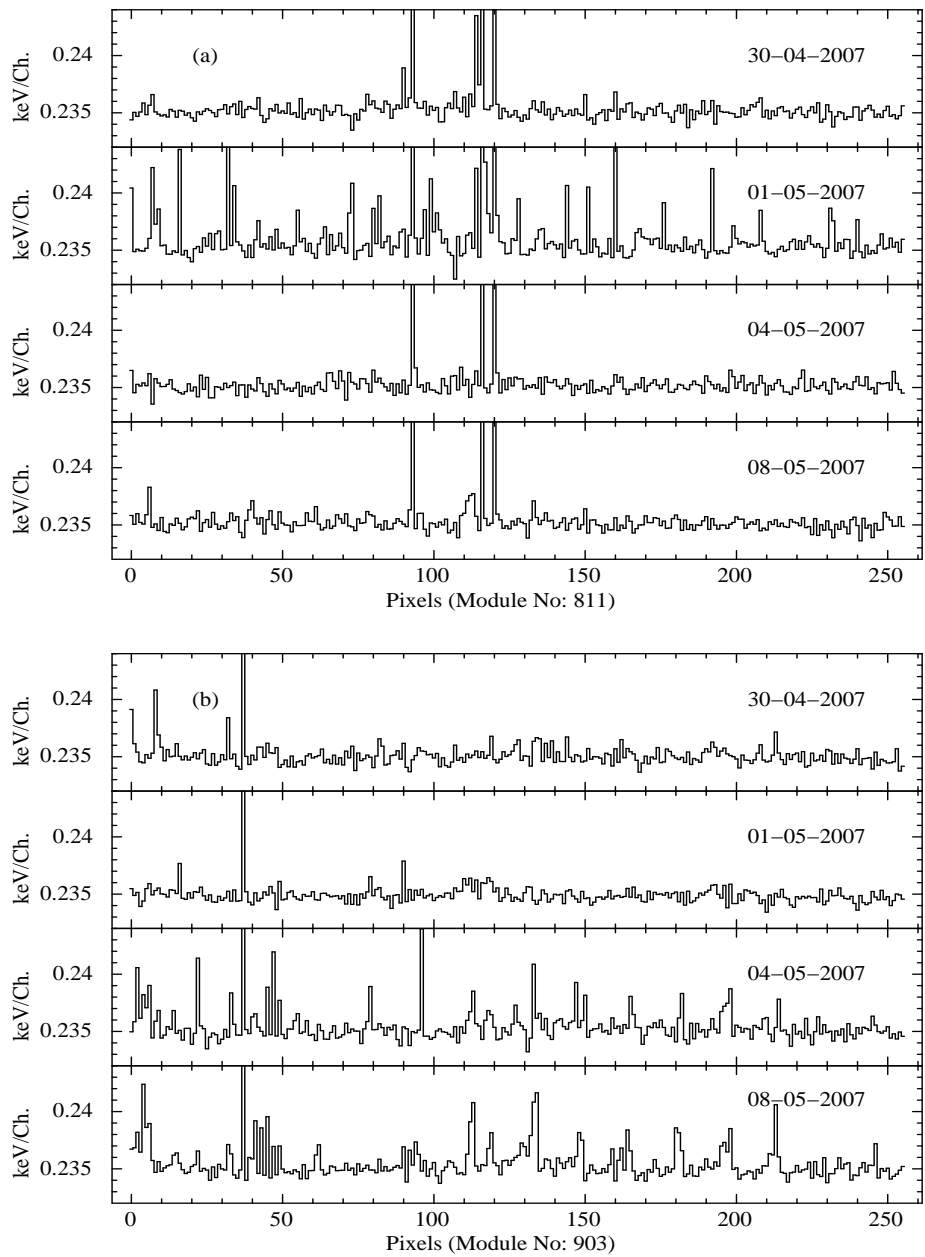

 \includegraphics[width=8cm, height = 12cm, angle = 270]{fig10a.ps}
\vskip 0.5 cm 
 \includegraphics[width=8cm, height = 12cm, angle = 270]{fig10b.ps}
\caption{Life test of two CZT modules depicting the gain (keV/Channel) variation.}
\end{figure}

\subsection{Life Test}

We performed a life test of CZT module to evaluate their response stability and 
behavior over a long time scale operation. The life test was carried out at room temperature. 
In Fig. 10 (a,b), we show gain variations of each pixel of two CZT modules monitored 
for a period of nine days. The gain of these modules remains almost constant over this time 
scale despite the pixels (every pixel has different gain and characteristics features) 
getting bad in between for a day and getting back to the normal after sometime which is 
basically due to temperature gradient effect on the modules. These effects were noticed in 
both the modules as seen clearly from the sudden rise in the gain of some channels (pixels). 
This effect is controllable with the help of cooling system to dissipate heat generated by 
ASIC. Similarly, a sudden rise in pixel count were found due 
to electromagnetic noise from the surrounding picked by the module, which makes them look like
hot pixels. On appropriate grounding of the power supply lines and the body of test setup, 
these pixels were found to be good.

Further parameters like offset and energy resolution were also under observation during the 
life test. The aim of life test was to measure the performance, device structural rigidity 
and stability with signal communication and power consumption are at their norms. On 
achieving the life test over the modules, there was no such major inconsistency noticed among 
the above parameters. This indicates that the CZT crystal properties remain invariant with 
time of detector operation. It could be concluded that the performance of OMS CZT modules 
(gold) are stable over a long duration run and hence can be used for space flight.

\subsection{CZT detectors for space-flight instrument}

Eight CZT modules (Gold quality) were selected for use in the RT-2/CZT. Out of them, 3 modules 
were selected for the flight payload on the basis of their performance. To qualify these 
modules for the space, thermal cycling tests were the first to be carried out for 10 cycles 
with temperature ranging from -35$^\circ$C to 65$^\circ$C with the transition rate of 
3$^\circ$C/min. After carrying out thermal cycling on these modules, there was not any drastic 
change found in the over all performance of the modules. After going through all the space 
qualification tests, these modules were subjected to life test which was carried out for 6 
days at room temperature. 
 
During the last tests, it was found that a couple of pixels became temporarily noisy. 
A few pixels were found permanently noisy or dead which were blocked from the respective 
modules. We found a 
single pixel which become noisy due to factory defect that is present in every single module, 
if the threshold is set below 40 keV. In Fig. 11, the results of both thermal cycling test and 
life test carried out on one of the flight module is plotted, in which gain of each pixel is 
monitored and compared with respect to each day performance. It was noticed that a uniform 
pixel gain was seen throughout the tests for all the modules, besides a few disqualified 
pixels. Similar results were obtained for the remaining two CZT modules for flight use. 

\begin{figure}[h]
\vskip -0.5cm
\centerline{
\includegraphics[width=12cm, height=8cm]{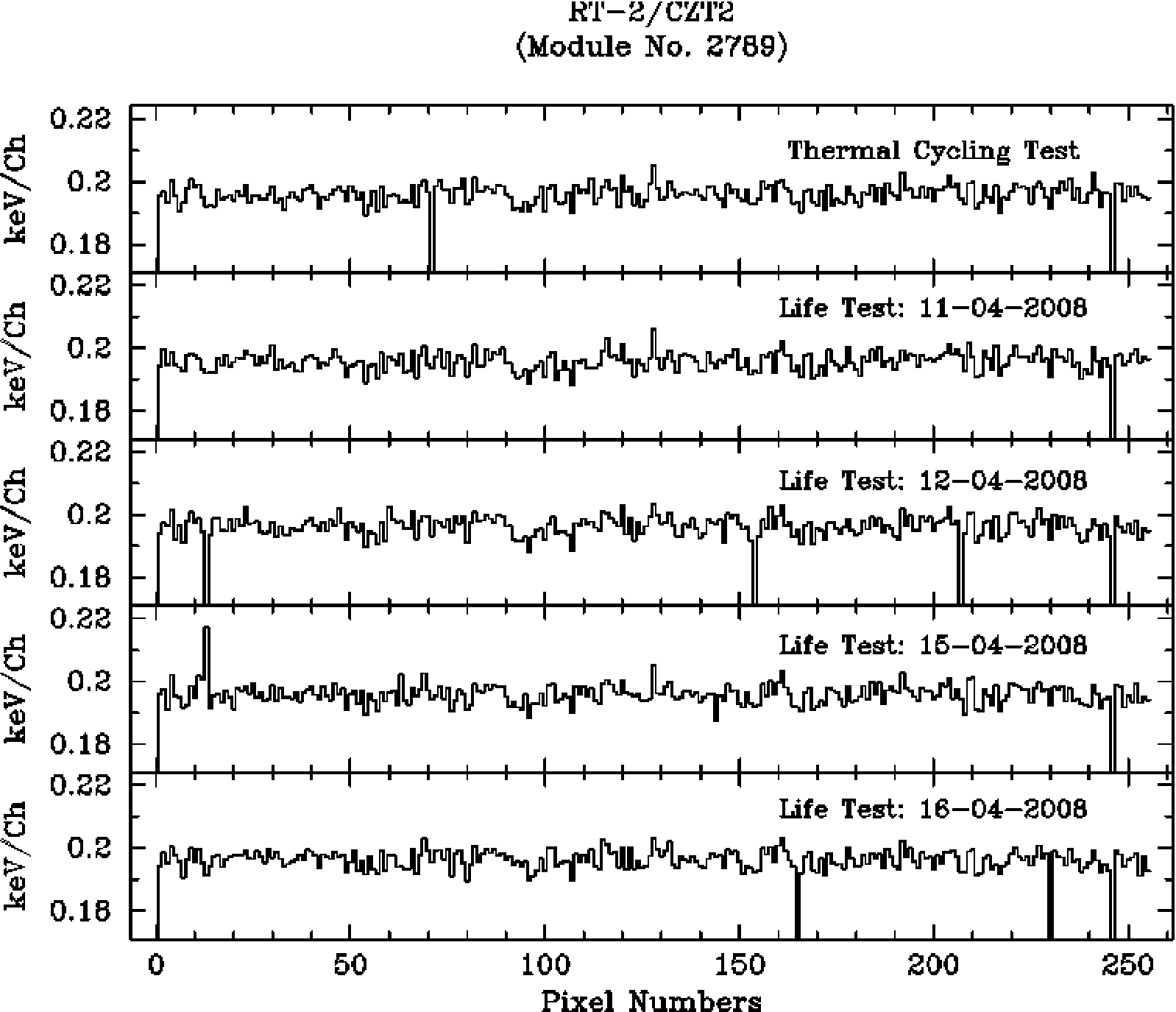}}
\caption{Thermal cycling and life test result of a CZT module depicting the variation of gain 
(keV/Channel) monitored for the period of 6 days.}
\label{}
\end{figure}

\section{The CMOS detector}

The RadEye1 CMOS (Fig. 12) is a large-area image sensor (procured from Rad-icon Imaging 
Corporation, USA). It is a complete detection system for high-resolution radiation imaging. 
The effective area of 24.6 mm x 49.2 mm is made up of a 512 x 1024 array of sensors, where 
each sensor module contains a two-dimensional photo-diode array with 48 $\mu$m pixel 
spacing. Each pixel has its own charge-to-voltage conversion. These CMOS detectors are 
basically visible imaging photo-diode detectors. But with the help of scintillator, these 
detectors can be used to detect X-rays and other energetic radiation. A Gd$_{2}$O$_{2}$S 
scintillator screen placed in direct contact with the photo-diode array, converts incident 
X-ray photons to light, which in turn is detected by the photo-diodes. RadEye1 CMOS imager 
provides a fully differential high-speed video signal, which is digitized with 12-bit 
resolution and transmitted to the processing electronics. Its operating temperature range is 
around 0$^\circ$C to 50$^\circ$C with dark current (noise) of $\sim$4000 electrons/sec at 
23$^\circ$C (room temperature), which gets approximately double at every 8$^\circ$C increase 
in temperature.

\begin{figure}[h]
\centerline{
\includegraphics[width=8cm, height= 3cm]{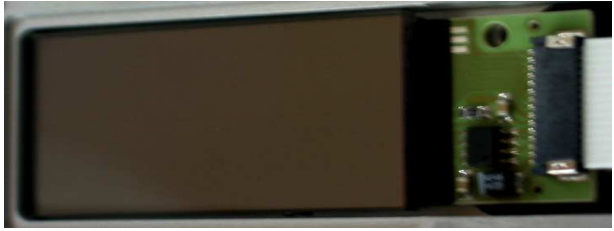}}
\vskip 0.1cm
\caption{RadEye1 CMOS detector.}
\label{}
\end{figure}

\subsection{Test Setup}

The implemented CMOS detector is basically a commercial product and as such it has no specific 
space qualification. Due to its high resolution imaging capability, we have decided to use 
CMOS as a detector onboard RT-2/CZT payload to perform a fine imaging of hard X-ray solar 
flares. So, the detector functionality was throughly verified at different temperatures before 
subjecting it to the space qualification tests.

The detector was enclosed in a plastic bag containing two small pouches of silica gel in order 
to prevent moisture condensation and with a thermistor mounted on the surface of the detector 
near to its window for monitoring the temperature. The entire package was made airtight by 
wrapping the bag with an adhesive tape. This setup was mounted on the thermoelectric cooler to 
cool the detector to the desired temperature with a variation of $\pm$ 1$^\circ$C from the 
set point. The whole assembly (the package plus thermoelectric cooler) was kept inside a 
cooled chamber which has the temperature operation range from 0$^\circ$C to 25$^\circ$C. The 
radioactive source Am$^{241}$ was mounted on a fixed position above the CMOS detector plane. 
The detector is connected to the test electronics (called Shadow-o-Box) using a connector of 
15 pins coming out of the sealed package. The Shadow-o-Box is powered with 5V supply from the 
adapter containing the processing electronics of the detector and it is connected to the input 
of the grabber card which is implemented in the computer based acquisition system. This card is 
acting as interface between the detector and system. 

\subsection{Test procedure}

The Shadow-o-Box is powered with 5V supply to make the CMOS detector operational. Using the 
system based application provided by the vendor, the data is retrieved from the detector to 
get a raw image of the events. This application needs only once a pixel map file and offset 
correction information along with exposure time before getting an image from the detector. We 
acquired the images of the radioactive source by keeping it at different distance above the 
detector surface with various exposure times (1s, 3s, and 6s). The temperature of the detector 
was varied from room temperature (23$^\circ$C) to 0$^\circ$C by step of 8$^\circ$C, in order 
to check its performance and noise characteristic as function of temperatures. 

\subsection{Results}

It was observed for $6.0$ sec of image acquisition, that there is a clear and distinct 
separation between background and source counts distribution. In Fig. 13, we have plotted 
source and background count distribution at different temperatures. An image of the 
radioactive source ($59.5$ keV emission peak) as detected by the CMOS is also shown in the 
same figure (right panel). On the other hand, for the $1.0$ sec data acquisition, the 
background variation is almost merged over the source spectrum and the image of the source 
is too faint and hardly visible. The overall results remain the same during 
the temperature variation from 23$^\circ$C to 0$^\circ$C. From the known count rate 
of the source, we estimate that an energy deposition of about $150$ keV per pixel 
(per sec) is required to clearly distinguish the source from the background, 
i.e., $150$ keV photon will produce approximately the counts of $15 - 20$ per 
pixel in comparison to the background counts $4 - 6$ per pixel.

\begin{figure}[h]
\vbox{
\vskip 0.0cm
\centerline{
  \includegraphics[width=9truecm,height=6truecm]{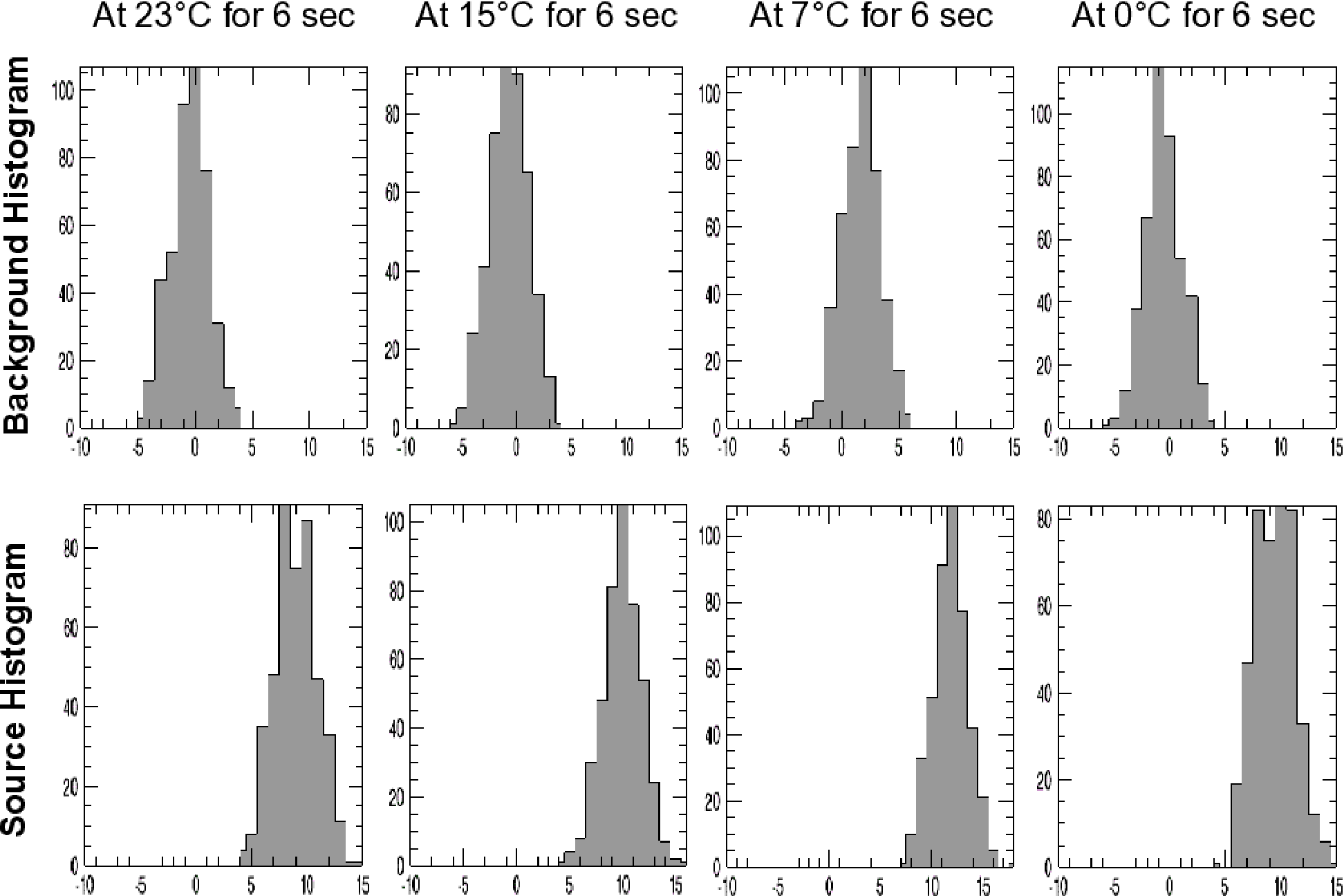}
\hskip 0.2cm
 \includegraphics[width=3truecm,height=5.5truecm]{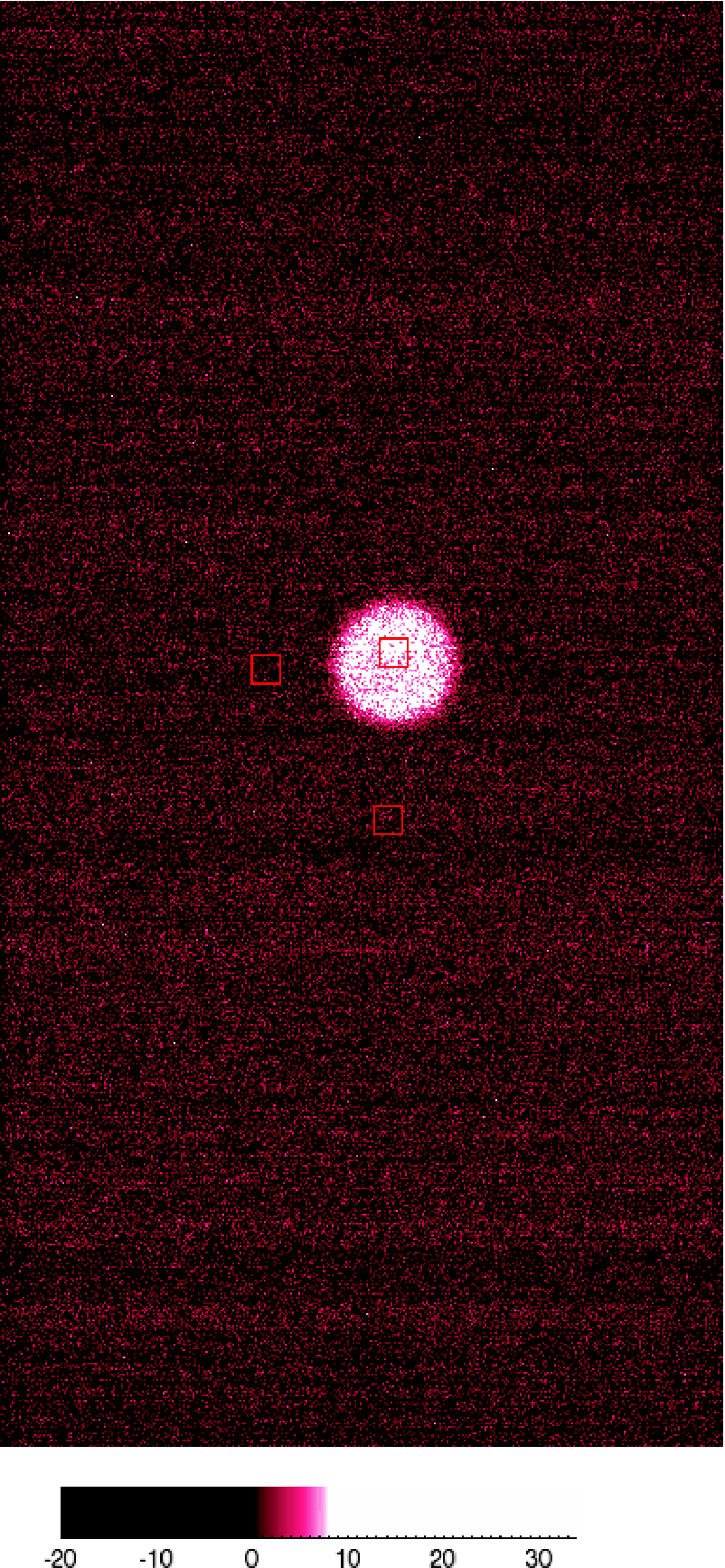}
}
\vspace{0.0cm}
\caption{Histogram of source and background variation at different temperature for $6$ sec 
image accumulation. In the right panel, an image of the Am$^{241}$ source is shown as 
detected by CMOS photo-diode at operating temperature of 23$^\circ$C.}}
\label{}
\end{figure}

\subsection {CMOS detector for the space-flight payload}

The CMOS detector went through all screening criteria of space qualification process used for 
the CZT detector modules: stabilization bake, thermal cycling and life test. During the tests, 
no significant change was found in the performance of CMOS. In Fig. 14, we have plotted the 
one bit background image (noise) as detected by the CMOS detector. The center dark line of the 
image is due to the manufacturing split of the entire 512 column into two halves. Detailed 
analysis on CMOS characterization with Fresnel Zone Plates (FZPs) mask is given in Nandi et al. 
(2010).

\begin{figure}[h]
\vbox{
\vskip 0.0cm
\centerline{
 \includegraphics[width=6.0cm,height=4cm]{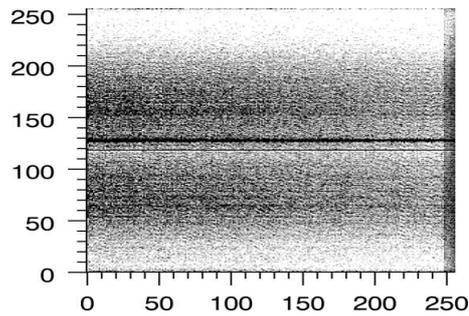}}
\vspace{0.0cm}
\caption{One bit background image (noise) as detected by CMOS}}
\label{}
\end{figure}

\section {Flight model testing of RT-2/CZT payload}

RT-2/CZT payload consists of 3 CZT modules and 1 CMOS detector. The overall detector 
specifications of RT-2/CZT payload are given in Table 3. Note that the geometric area of CMOS 
is restricted to 25 mm $\times$ 25 mm and neighboring pixels in a 2 $\times$ 2 matrix are 
clubbed together to increase the pixel dimension to 100 $\mu$ $\times$ 100 $\mu$ and 
correspondingly reducing the number of pixels to 256 $\times$ 256.

\begin{table}[h]
\centering
\centerline {Table 3: Specifications of RT-2/CZT payload}
\begin{tabular}{|c|c|c|}
\hline
Detector type & CZT & Gd$_{2}$O$_{2}$S +  CMOS \\ 
\hline
Thickness (mm) & 5 & 3 \\ 
\hline
Size (mm) & 40 X 40 (3 Numbers) & 25 X 25 \\ 
\hline
Read out (pixel) & 256 x 3 & 256 X 256 \\ 
\hline
Geometric area (cm$^{2}$) & 48 & 6.3 \\ 
\hline
Energy resolution ($@59.9$ keV) & 10$\%$ & NIL \\ 
\hline
Energy range & 20 - 150 keV & 25 - 150 keV \\ 
\hline
Time resolution & 1 s counts and 100 s imaging \& spectrum & 100 s imaging \\
\hline
\end{tabular}
\end{table}

\subsection{RT-2/CZT Electronics}

The RT-2/CZT detector assembly contains a Main board (MB) with 3 CZT Modules and a CMOS 
module. CZT modules are connected on SPI bus which is further connected to Front End Board 
(FEB). The Main board (MB) is connected to FEB to interface various signals. The other 
interface of MB is with HV supply, that provides the appropriate bias to CZT modules. The 
high voltage -600 V bias is provided through a circular connector to each CZT module. The 
schematic block diagram of RT-2/CZT is shown in Fig. 15.

\begin{figure}[h]
\centerline{
 \includegraphics [width = 8cm, height = 6cm]{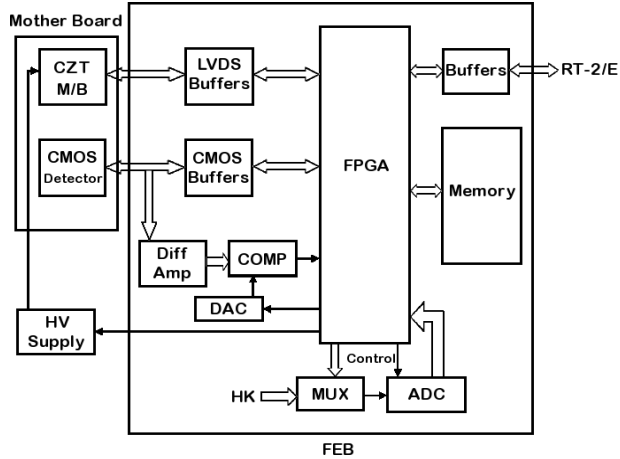}}
\vskip 0.1 cm
\caption{The schematic block diagram of RT-2/CZT.}
\label{}
\end{figure}

The FEB is the most essential part of the entire electronics. Besides having interfaces with 
the Motherboard, it has interfaces with the processing unit (RT-2/E). The FEB contains some 
common interface circuits. One component is the low noise differential amplifier with 50 MHz 
bandwidths which is employed to amplify the CMOS detector video output. This amplified output 
is compared with the CMOS threshold given by FPGA to produce one bit image. FPGA is the heart 
of this board which controls all operations and acquires data from detectors. The SPI 
interface to CZT module is handled by logic inside FPGA. It is interfaced with CZT modules 
through LVDS buffers. CZT detector data acquisition and control is done by FPGA through this 
SPI interface. 

The FEB process all the commands needed to configure the CZT modules of operations. The main 
function of this board is to acquire data from both the detectors and to form spectrum and image 
in memory. It has a ping-pong memory bank. It sends acquired data to RT-2/E after receiving 
`data-send' command from RT-2/E. The interface circuitry is necessary to read-out CMOS 
detector video data which is in differential voltage form. DAC converts a digital threshold 
value provided by FPGA to analog value. Output of the comparator is considered to be one bit in FPGA 
for each pixel. Differential signal conditioning offers many advantages over single ended 
technologies. LVDS signal conditioning centers on 1.25 V with a 350 mV swing and does not 
depend on power supply voltage. As CZT modules are having LVDS lines, 
they are interfaced to FPGA in FEB using LVDS buffers.

The RT-2/CZT payload draws raw input power (from satellite power bus during flight) from 
external supply of 27$^{+7}_{-3}$ Volt. The total power consumption is limited to 7.5 Watt. 
The input power is converted to $\pm$ 15 V and +5 V with the help of the MDI unit for the 
required supply of the detector and front-end electronics. A non-controllable high voltage 
generator (made by PICO) is used to bias (fixed) the CZT detector with -600 Volt. The 5 Volt 
supply to CMOS is derived from +15 V. The operation of RT-2/CZT payload is completely 
commandable and controlled by the processing electronic device RT-2/E (Sreekumar et al. 2010).

\subsection{RT-2/CZT operational mode}

There are two different modes in which RT-2/CZT can be operated: the Event Mode and the Normal 
Mode. Each event registered by RT-2/CZT is characterized by 2 words. The basic data structure 
is given below (refer to Sreekumar et al. 2010, for details).

\begin{itemize}
\item {\bf CZT Event mode data format}

The CZT Event mode consists of 32 bits of data words (referred to as bit numbers D0 through 
D31) with the following configuration:

\begin{enumerate}
\item D1 - D0 : Detector ID. 0 to 2 for 3 CZT modules.
\item D9 - D2 : Pixel ID (0 to 255).
\item D19 - D10 : ADC value of the detected signal.
\item D31 - D20 : Time (with a resolution of 0.3 msec).
\end{enumerate}
\end{itemize}
The data stored in memory is sent to RT-2/E unit each second. The maximum numbers of events 
that can be accumulated are 4032.

For the normal mode, data structure for CZT and CMOS detector is given below:

\begin{itemize}
\item {\bf CZT Normal mode data format}

In the normal mode, CZT spectral and image data are accumulated every second and count rates 
are accumulated every 10 ms. A total of 5832 words (each word data consists of 16 bits) of 
memory space are allocated for CZT data.

Image data are accumulated in 4 energy bands or channels (approximately equally spaced in the 
20 - 150 keV range, though the channel boundaries can be changed by command) leading to 12 
images of 256 pixels (for the 3 modules). Apart from the image, four channel information i.e. 
counts are stored separately for 10 ms. Therefore, a total of 12 counters (3 modules x 4 
channels) are stored for every 10 ms. Spectral data of each module is accumulated in 512 
spectral channels. 

\begin{enumerate}
\item Image block (3072 words): 1 K words per CZT, 4 channel X 256 pixels X 1 word.
\item Spectrum block (1536 words):  512 words per CZT
\item Timing blocks (1200 words): 3 CZT detector X 100 timing words x 4 channels x 1 word 
(counters in each block will count for 10 ms).
\item Counter block (24 words): 12 counters (2 words each)
\item VCO block (1 word): 2 bytes
\item Special words (8 words): Satellite telemetry word, temperature, Command sent, Data 
read against command, event number, CMOS line number, Calibration result identification 
word and Calibration status.
\end{enumerate}
\end{itemize}

\begin{itemize}
 \item {\bf CMOS data format}
\begin{enumerate}
\item Image block (4096 words): 256 x 256 pixels, 1-bit image.
\item Sum (512 words): Vertical sum (256 words) + Horizontal sum (256 words)
\end{enumerate}
\end{itemize}

\subsection{Test and Evaluation of RT-2/CZT payload}

Three CZT and one CMOS detector were selected for use in the flight RT-2/CZT payload based 
on performance obtained during the space screening test and satisfactory results during test
and evaluation of the Qualification Model. The Qualification Model is a pre-flight payload and 
its test results are not summarized in this paper. Only results with the RT-2/CZT Flight 
payload are discussed in the following sections.

\subsection{Test setup}

In the flight condition, RT-2/CZT would be fully controlled by RT-2/E. The overall testing of 
all 3 payloads (RT-2/S, RT-2/G \& RT-2/CZT) with RT-2/E and Ground Check-out system is 
discussed in Sreekumar et al. (2010). We have tested and verified the functionality of 
RT-2/CZT payload independently, with a computer through an isolator unit, called the OPTO 
device, SCB-68 connector box and NI data acquisition card (PCI 6534). The OPTO device along 
with the read-out software were developed using NI LabVIEW platform. The OPTO device is used 
to isolate the payload electrically from the computer using buffers and opto-isolators. The 
test setup block diagram for qualifying the RT-2/CZT payload is shown in Fig. 16. 

\begin{figure}[h]
\centerline{
 \includegraphics [width = 8cm, height = 6cm]{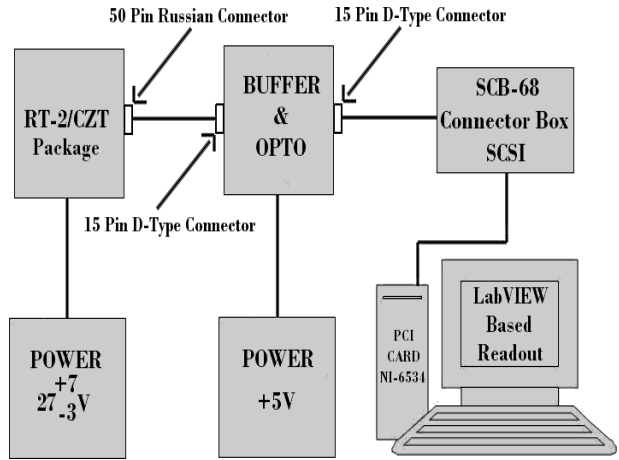}}
\vskip 0.1 cm
\caption{The block diagram of the test set-up used for testing of RT-2/CZT payload.}
\label{}
\end{figure}

In the laboratory, RT-2/CZT detectors were calibrated with two radioactive sources Am$^{241}$ 
(59.5 keV) and Cd$^{109}$ (88.0 keV). The calibration results, health condition of the payload 
and overall detector functionality are discussed in the following sections. The operation of 
RT-2/CZT payload in different modes, channel boundary change of timing data, HV control etc. 
are commandable. The command structure of RT-2/CZT operation is given in Table 4.

\subsection{Flight model test results}

CZT detectors are powered with supply voltage of 27 Volt for operation and high voltage supply 
of -600 Volt for setting the threshold, whereas CMOS is operated with normal 5 Volt supply. 
The overall power consumption of the payload is 6.75 Watt.

The RT-2/CZT Flight payload contains three CZT detectors with following serial numbers and 
calibrated energy channel information as given below:

\begin{itemize}
\item CZT-1 (Serial No - 2783) - (116 channel corresponds to 20 keV)
\item CZT-2 (Serial No - 2789) - (115 channel corresponds to 20 keV)
\item CZT-3 (Serial No - 2954) - (114 channel corresponds to 20 keV)
\end{itemize}

Health informations of RT-2/CZT payload are fed to processing device RT-2/E through ADC. Eight 
channel ADC output are shown in the table 5.

\begin{table}[h]
\centering
\centerline {Table 5: Health informations of RT-2/CZT payload}
\begin{tabular}{ccccc}
\hline
\textbf{Channel No.} & \textbf{Description} & \textbf{Operating voltage level} \\
\hline
0 & Supply Voltage & 5.0 $\pm$ 0.5 V \\
1 & Thermistor & 1.5 - 5.5 V \\
2 & GND & 0 V \\
3 & GND & 0 V \\
4 & HV OFF/ON & 0.0 V/5.0 V \\
5 & CZT Supply (DVDD) & 3.6 V \\
6 & CMOS Supply & 5.0 V \\
7 & FPGA Supply (VCCA) & 2.5 V \\
\hline
\end{tabular} 
\end{table}

\subsubsection {Spectrum and image validation of CZT modules}

All 3 CZT modules (CZT-1, CZT-2, CZT-3) of RT-2/CZT payload have image and spectral 
information. Apart from the background image and spectrum, we have exposed the modules with 
a radioactive source Am$^{241}$ for 20 sec. Each module has detected the emission peak of 59.5 
keV along with the calibration source (Co$^{57}$) peak of 122 keV. The 1024 channels source 
spectrum and 256 pixels image of all 3 CZT modules are shown in Fig. 17. The measured peak 
channel number versus the emission peaks energy is given in Table 6.

\begin{table}[h]
\centering
\centerline {Table 6: Peak channel informations of CZT modules}
\begin{tabular}{lcccc}
\hline
{\bf Source} & {\bf CZT-1} & {\bf CZT-2} & {\bf CZT-3} \\ 
(Energy in keV)	 & (Channel no.) & (Channel no.) & (Channel no.) \\ 
\hline
{\bf Am$^{241}$ (59.5)} & 311.9$^{+0.6}_{-0.8}$ & 319.2$^{+0.5}_{-0.3}$ & 309.5$^{+0.7}_{-0.7}$ \\ 
{\bf Co$^{57}$ (122.0)} & 630.6$^{+2.2}_{-2.4}$ & 638.7$^{+2.6}_{-2.7}$ & 628.4$^{+3.0}_{-3.5}$ \\ 
\hline
\end{tabular}
\end{table}

\begin{figure}[h]
\vbox{
\vskip 0.0cm
\centerline{
   \includegraphics [width=5cm, height=5cm]{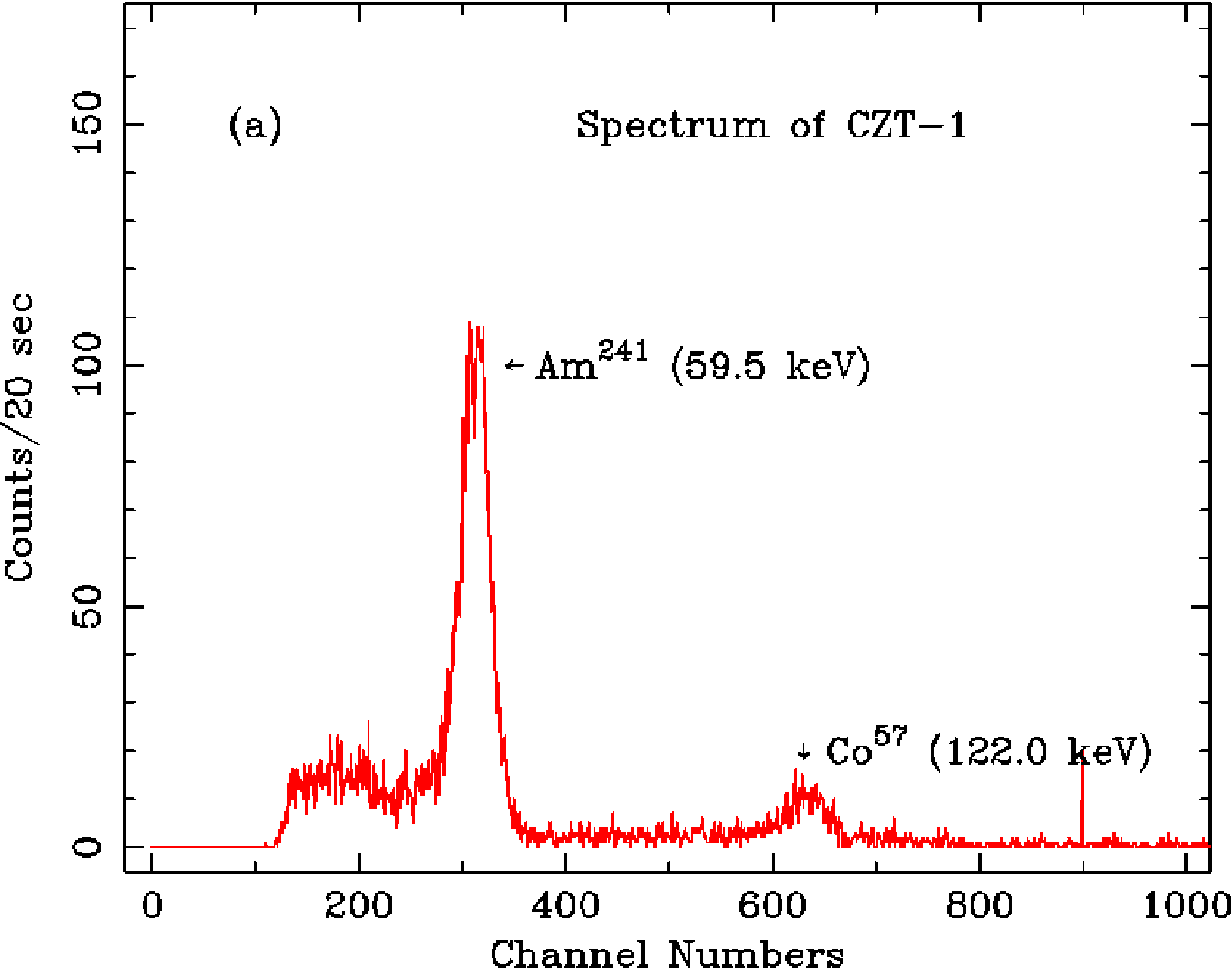}\hskip 0.2cm
   \includegraphics [width=5cm, height=5cm]{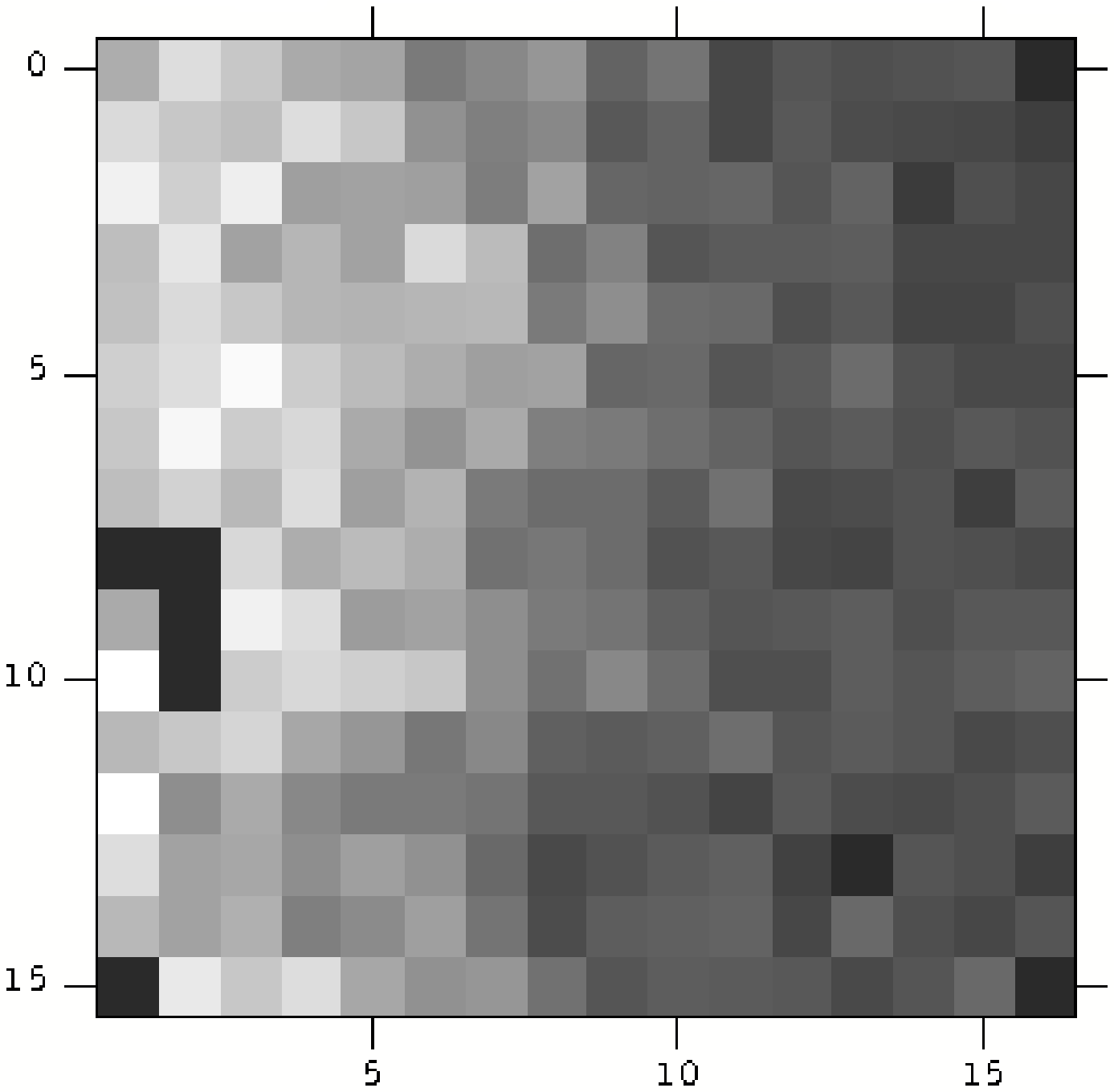}}}
\vskip 0.1cm
\vbox{
\centerline{
\includegraphics [width=5cm, height=5cm]{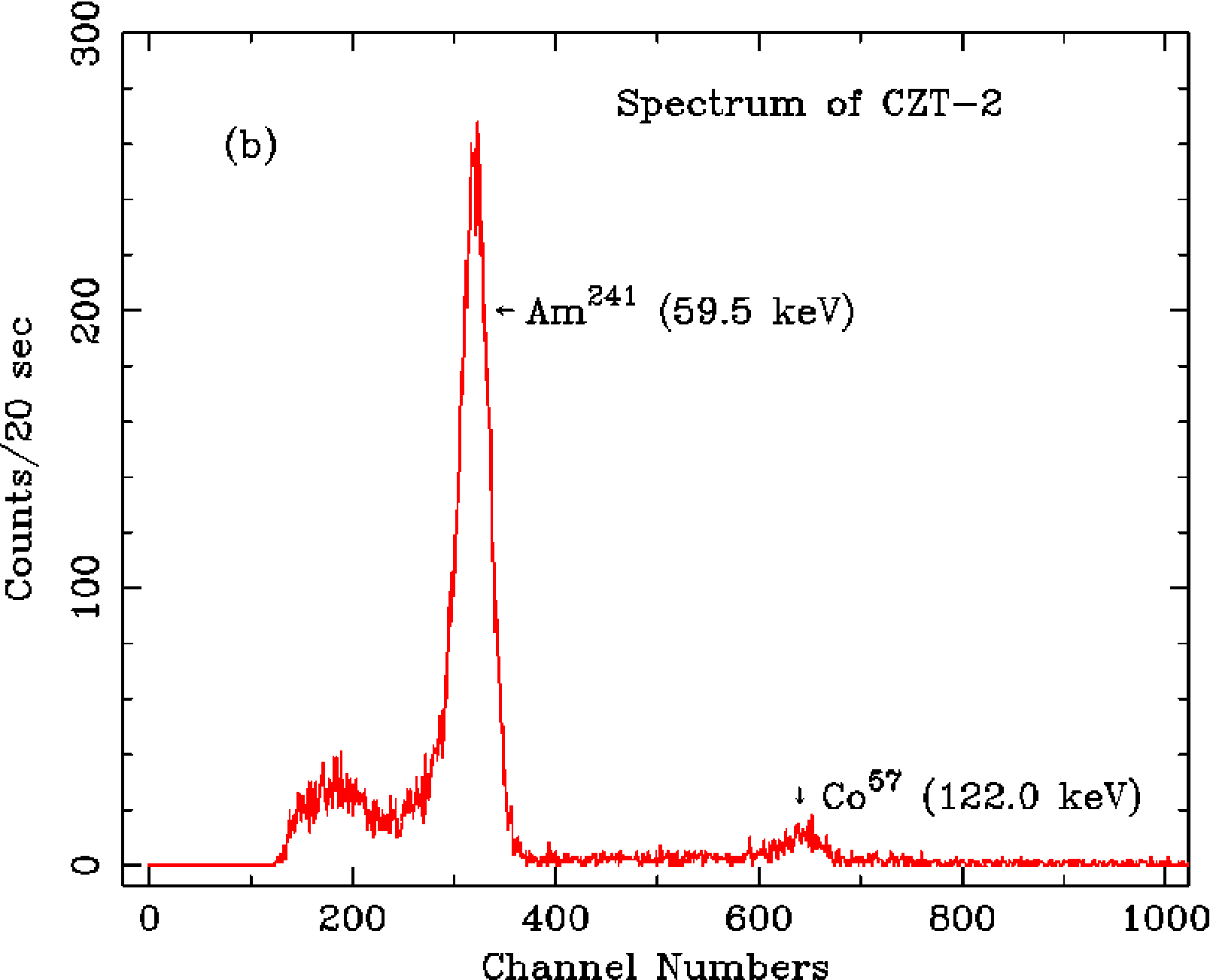}\hskip 0.2cm
\includegraphics [width=5cm, height=5cm]{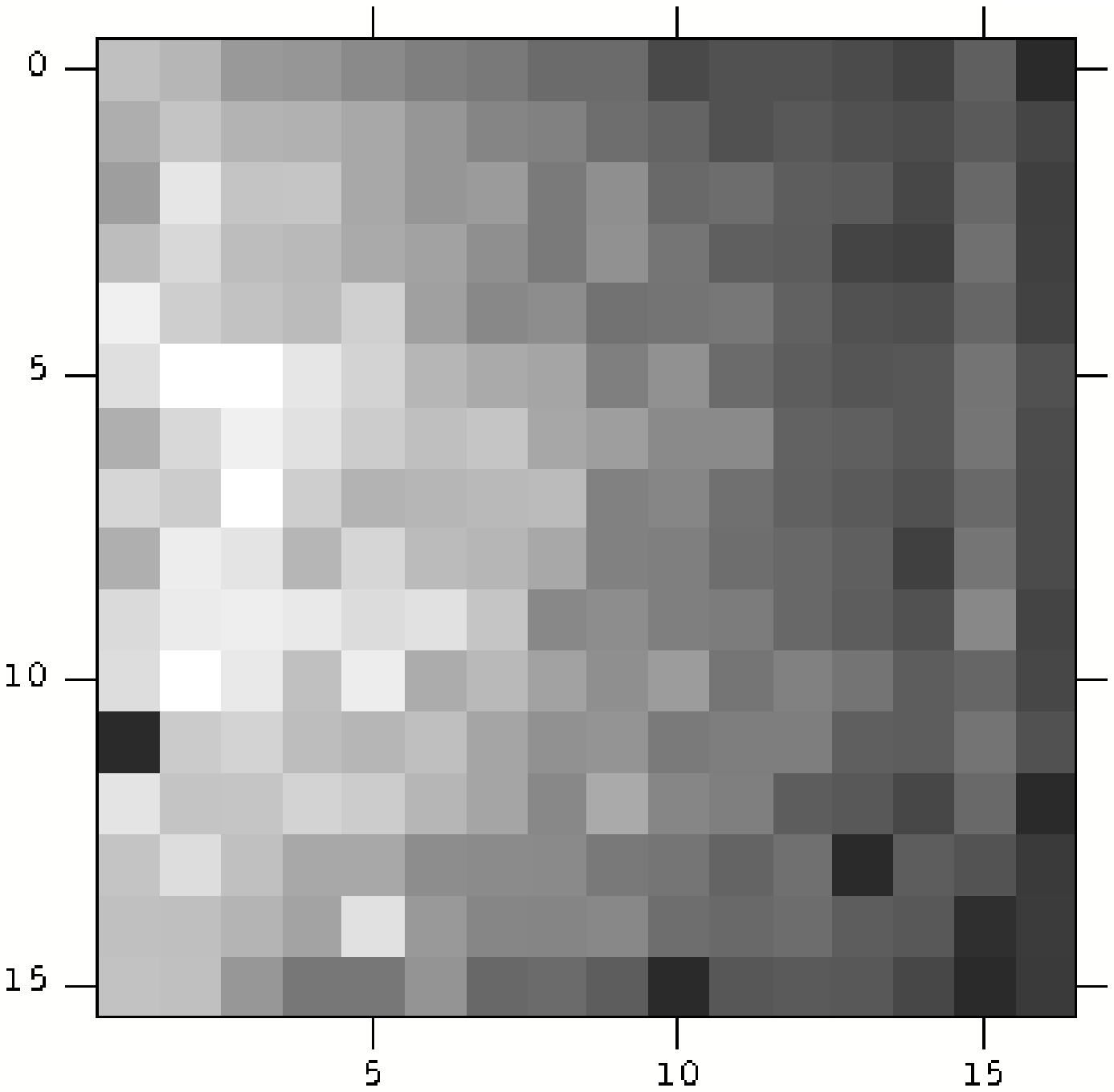}}}
\vskip 0.1cm
\vbox{
\centerline{
\includegraphics [width=5cm, height=5cm]{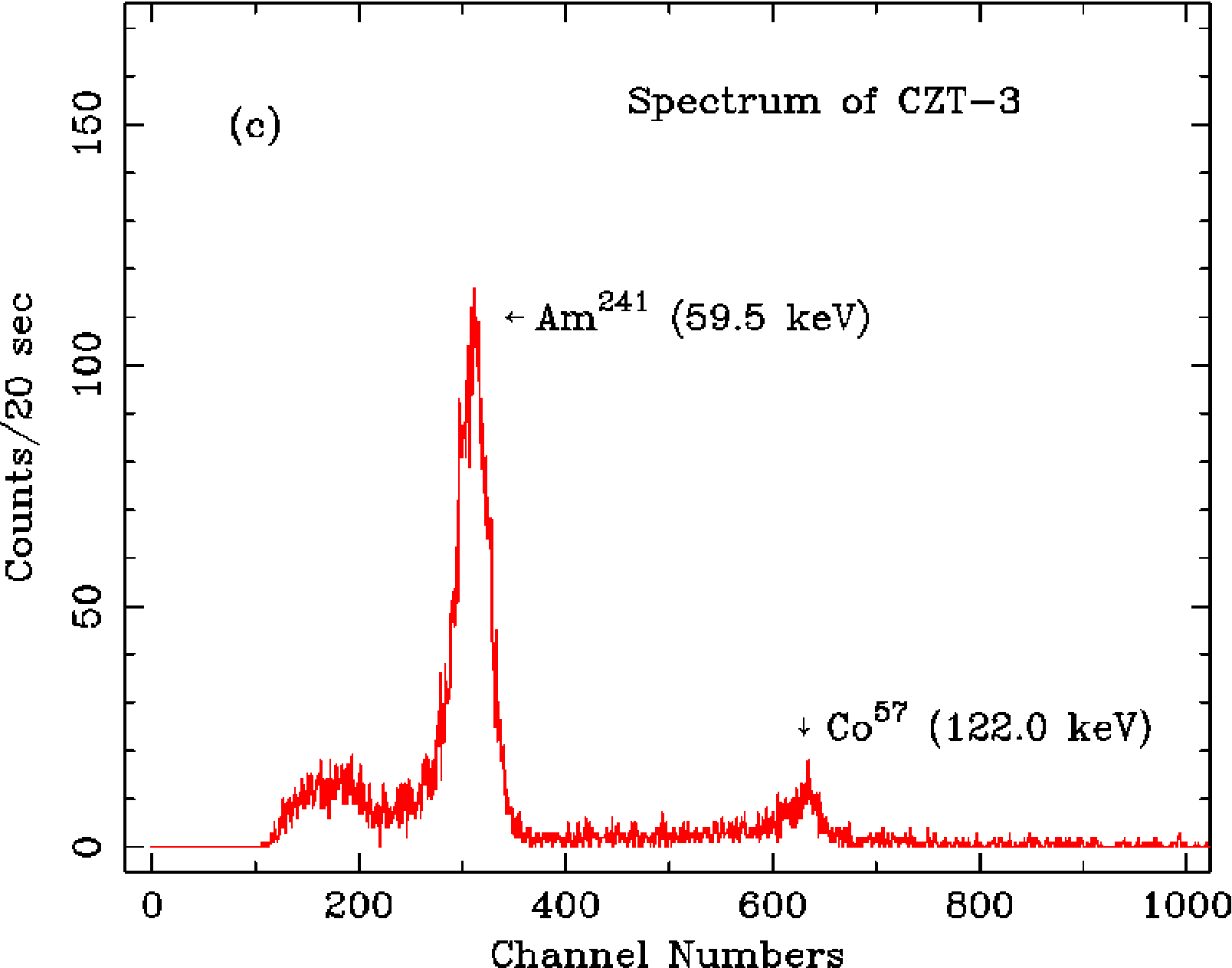}\hskip 0.2cm
\includegraphics [width=5cm, height=5cm]{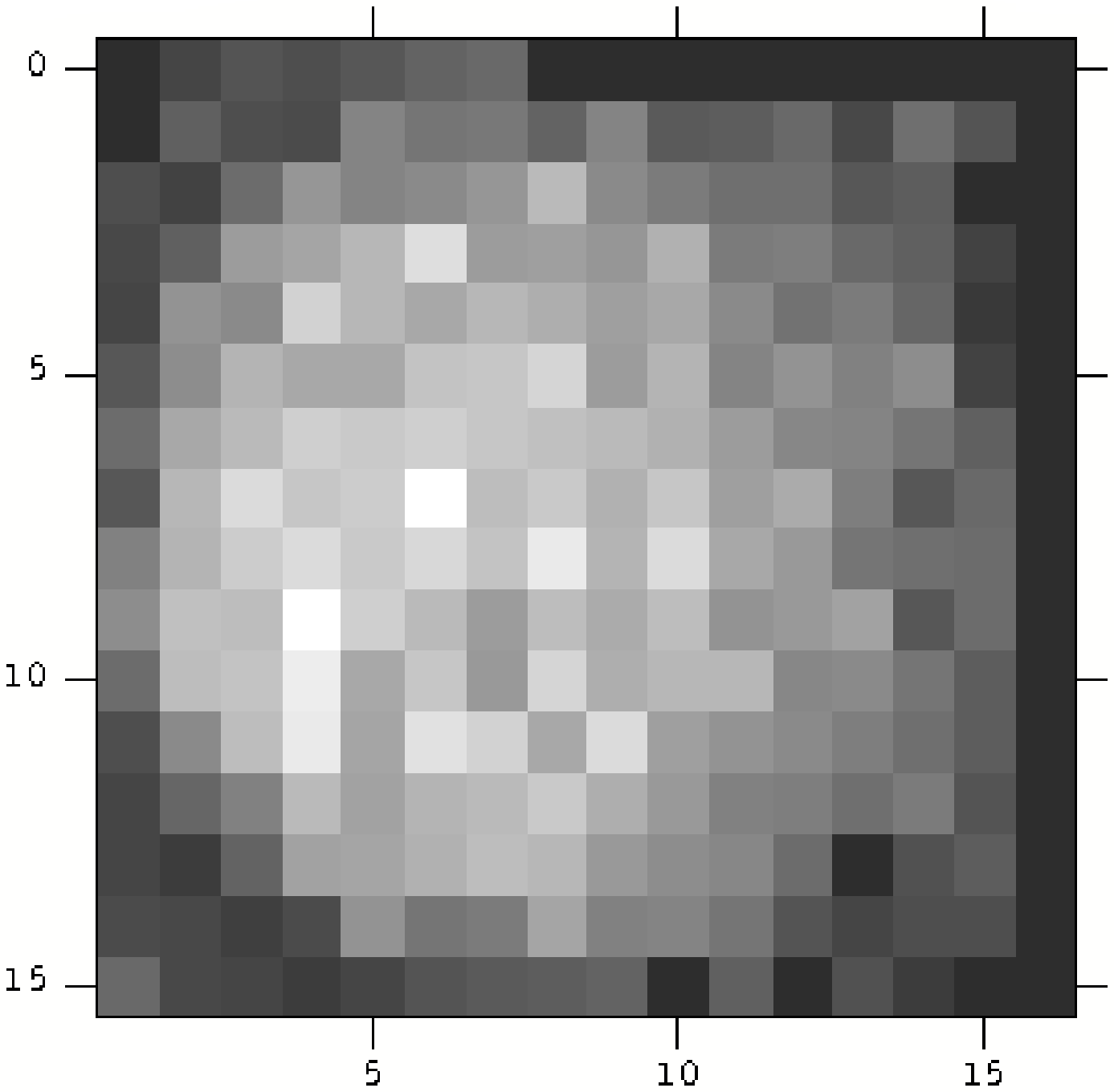}}}
\vspace{0.0cm}
\caption{Spectrum and image of three CZT modules of Flight payload, (a) CZT-1, (b) CZT-2 \& 
(c) CZT-3. All modules are irradiated with radio-active source Am$^{241}$ and Cd$^{109}$. Each 
spectrum is sum counts of all the pixels (256 pixels) of each detector module.}
\label{}
\end{figure}

Image of CZT-1 and CZT-2 module show that a few pixels (left side of the module) were 
illuminated with the source and not all the 256 pixels. This is due to the fact that source is not 
properly placed. On the other hand, image of CZT-3 is 
fully illuminated and pattern is in circular form. As per design (see Fig. 2 of Nandi et al. 
2010), we have used Coded Aperture Mask (CAM) in CZT-1 and CZT-2 and Fresnel Zone Plate (FZP) 
in CZT-3 as a coder to cast image on the detector plane (Nandi et al. 2010). It is noted that 
during testing a few pixels were found to be noisy, which are dark pixels (shown in the image).

\subsubsection {Channel-Energy calibration of 3 CZT modules}

Multiple radioactive sources were shinned over all the 3 CZT modules to calibrate the module 
and to find energy resolution at 10$^\circ$C. For calibration we have used three radioactive 
sources: Am$^{241}$ (59.5 keV), Cd$^{109}$ (22.0 keV and 88.0 keV) and Co$^{57}$ (122.0 keV). 
The gain calibration is applied to each pixel of a module during the ground testing of the flight
payload. This is done by feeding the input of the calibration file generated using IDL code
(discussed in section 2.3) from the test data of flight CZT modules. The final spectrum is an
integrated one of all the pixels of a CZT module.
In Fig. 18, the spectrum of the CZT-1 module with four distinct emission peaks are shown. The 
corresponding channel and energy values of the emission peaks are used for channel-energy 
calibration in the energy range of 20 - 150 keV. The same procedure is repeated for the other 
two modules. We find that the rms deviation from a linear fit of all three modules in the 
energy range between 20 to 150 keV is $<$0.5\%. From the above analysis, we have calculated the 
gain (keV/Ch), Offset (keV) and energy resolution at 59.5 keV and 122.0 keV of all 3 CZT modules. 
The results are summarized in Table 7.

\begin{table}[h]
\centering
\centerline {Table 7: Calibration specifications of CZT modules}
\begin{tabular}{lccccc}
\hline
\bf Module & \bf Gain & \bf Offset & \bf Energy res. (\%) & \bf Energy res. (\%) \\
	 & (keV/Ch) & (keV) & (@59.5 keV) & (@122.0 keV) \\ 
\hline
CZT-1 & 0.201$^{+0.004}_{-0.004}$ & 3.43$^{+1.44}_{-1.49}$ & 11.02$^{+0.41}_{-0.39}$  & 5.74$^{+0.39}_{-0.22}$ \\ 
CZT-2 & 0.199$^{+0.004}_{-0.001}$ & 2.90$^{+0.25}_{-0.25}$ & 09.89$^{+0.22}_{-0.11}$  & 5.28$^{+0.28}_{-0.20}$ \\ 
CZT-3 & 0.204$^{+0.003}_{-0.004}$ & 3.21$^{+1.28}_{-1.32}$ & 11.99$^{+0.37}_{-0.30}$  & 6.32$^{+0.47}_{-0.24}$ \\ 
\hline
\end{tabular}
\end{table}

\begin{figure}[h]
\centering
 \includegraphics [width = 6.0cm, height = 12cm, angle=270]{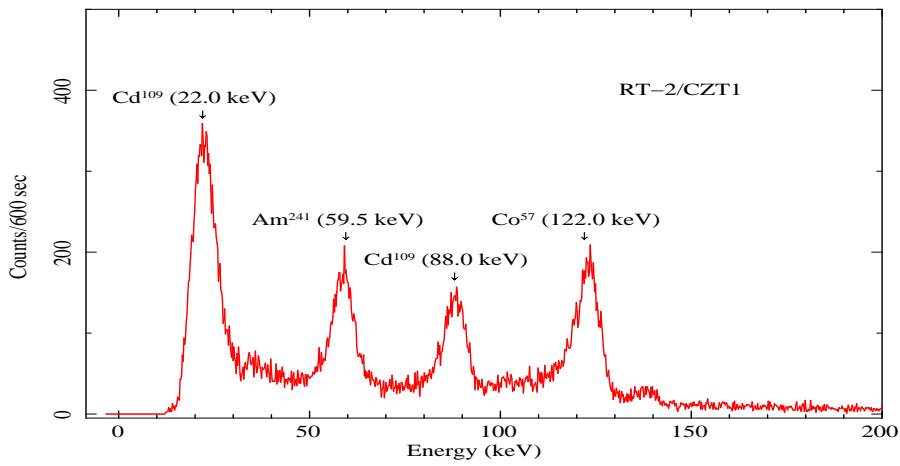}
\vskip 1.0 cm
\caption{Spectrum of CZT-1 module with 4 emission peaks of radioactive sources. Energy of emission 
peaks are marked.}
\label{}
\end{figure}

\subsubsection {Temperature effect on Pulse Height (PH) \& Energy resolution of CZT-2}

\begin{figure}[h]
\centering
 \includegraphics [width = 6.0cm, height = 12cm, angle=270]{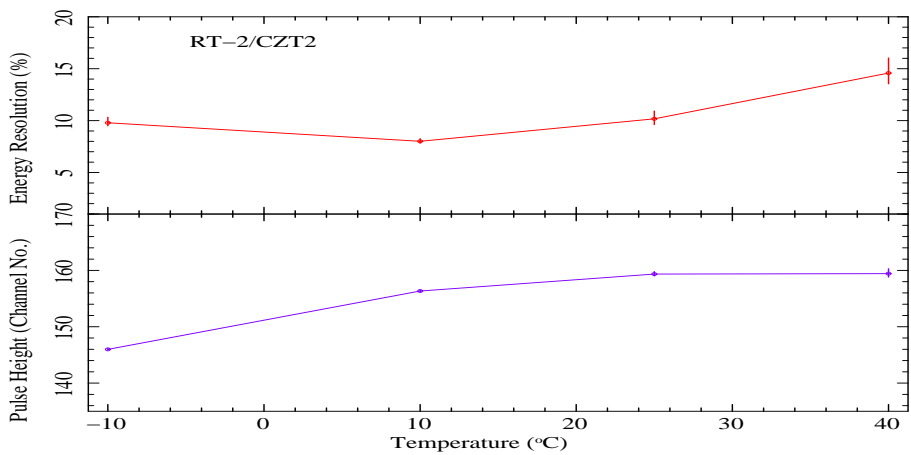}
\vskip 0.3 cm
\caption{PH and Energy resolution variation of CZT-2 module with temperature. Error values are
indicated on each data points.}
\label{}
\end{figure}

We made a systematic study of Pulse Height (PH) variation of one of the CZT module (CZT-2) 
with temperature. Payload temperature was varied from -10$^\circ$C to 40$^\circ$C. Emission 
peak of Am$^{241}$ (@59.5 keV) is calibrated as PH and its variation is noted at different 
temperatures. PH variation with temperature is plotted in Fig. 19 (bottom panel) and it is 
noted that maximum variation of $\sim$6.4\% is observed from the normal operating temperature 
at 10$^\circ$C.

In the top panel of Fig. 19, we have plotted the variation of energy resolution (@59.5 keV) 
at different temperatures. It is noted that the best resolution of $\sim$8.5\% is achieved in 
the temperature range of 10$^\circ$C to 20$^\circ$C and resolution become worse at 40$^\circ$C.

\begin{table}[h]
\renewcommand{\arraystretch}{1.3}
\centering{Table-4: RT-2/CZT control commands}
\centering
\begin{tabular}{|c|c|c|}
\hline
{\bf Command Type} & {\bf Command} & {\bf Description} \\
\hline
Range setting & 0x40xx & Set range 0 value : xx*2\\
\cline{2-3}
(8 bit value) & 0x41xx & Set range 1 value : xx*2\\
\cline{2-3}
 & 0x42xx & Set range 2 value : xx*2\\
\hline
CMOS start line & 0x43xx & Set the start line of CMOS frame : xx + 256\\
\hline
CMOS upper threshold & 0x44xx & Set the CMOS threshold : xx*16\\
\hline
CZT constant & 0x45xx & Set CZT constant, Constant = xx/8\\
\hline
Event mode disable & 0x460x & Disable CZT from event mode\\
and & & D0: CZT0, D1: CZT1, D2: CZT2\\
threshold & & (0:disable; 1:enable)\\
& & D4-D3: CZT select for satellite telemetry counts\\
\hline
RT-2/CZT HV setting & 0x47xx & Set HV value for RT-2/CZT from DAC\\
 & & xx=00; HV OFF, xx=01: HV ON\\
\hline
One second command & 0x4800 & Set RT-2/CZT in normal mode\\
\cline{2-3}
 & 0x4801 & Set RT-2/CZT in test mode\\
\cline{2-3}
 & 0x4803 & Set RT-2/CZT in test mode \& EEPROM data verify\\
\hline
Frame interval & 0x49xx & Set the time interval in which\\
 & & frames of CMOS are integrated\\
\hline
CZT parameters & 0x4Axx & D5: parity (0)\\
 & & D4: 3.3V (0: ON, 1: OFF)\\
 & & D3: 1.5V (0: ON, 1: OFF)\\
 & & D2: Reset CZT (0: Release, 1: Reset)\\
 & & D1-D0: CZT number\\
 & & (00 - CZT0, 01 - CZT1, 10 - CZT2)\\
\hline
CMOS constant & 0x4Bxx & Set CMOS constant, Constant = xx*16\\
\hline
Memory select & 0x4C0x & D2: RAM section select\\
 & & D1, D0 : EPROM page select\\
\hline
Vertical sum & 0x4Dxx & Set vertical sum threshold of CMOS,\\
threshold & & Threshold = xx*8\\
\hline
CMOS control & 0x4Exx & D7: CMOS control\\
\& clock & & D6 - D0: Clock 20M/(2*(xx+1))\\
\hline
CMOS threshold & 0x4Fxx & D7: calibration (1: For new threshold)\\
calibration & & (0: add offset to the present average\\ 
 & & value of 32 frame Data\\
 & & D6 - D0: threshold for calibration\\
\hline
EPROM address set & 0x5xx & Address : xxx*2\\
(12 bit value) & & \\
\hline
CZT command mode & 0x6000 & Set CZT module in command mode\\
\hline
CZT event mode & 0x6001 & Set CZT module in event mode\\
\hline
Reset RT-2/CZT & 0x6100 & Reset RT-2/CZT\\
\hline
RT-2/CZT start & 0x6101 & Start RT-2/CZT\\
\hline
EEPROM data & 0x62xx & Write data in EEPROM\\
\hline
EEPROM unlock & 0x6303 & \\
\hline
EEPROM lock & 0x6383 & \\
\hline
CZT commands & 0x7xxx & 7xxx - 7 00zz xx\\
 & & zz: CZT number xx: command\\
 & & 7 1xxx xx data write\\
 & & 7 0100 00 data read\\
\hline
\end{tabular}
\end{table}

\section{Conclusion}

We have presented the test results of several CZT detector modules along with the 
performance of the 3 selected CZT modules for the RT-2/CZT payload. During the room 
temperature tests, we observed varied pixels behaviors in terms of response and 
performance. The numbers of disqualified pixels also vary from module to module. 
We recognized two different types of bad modules whose behavior can be understood either 
on the basis of poor CZT crystal or bad ASIC. CZT module functionality appears to vary with 
temperature. This was revealed by both cold test and also during the space qualification 
screening procedure. Though the CZT modules can operate at room condition, their best average 
energy resolution is achieved in the temperature range between 10$^{\circ}$C to 20$^{\circ}$C. 
Beyond these temperature ranges, some of the pixels may become noisy showing a large variation 
in counting rate. The average energy resolution which varies with temperature is not uniform 
for all the modules. This implies that any selection of good CZT detector module for the 
flight payload shall be based on a trade off between the best average energy resolution and 
number of disqualified pixels. From the results obtained at 15$^{\circ}$C, 
we concluded that the number of disqualified pixels should not be more than 5.
The results of various flight 
screening tests conducted over the RT-2/CZT flight payload shows that the overall performance 
of CZT modules and CMOS detector remains invariant even after going through various 
environment conditions along with flight electronics. As expected, the flight CZT modules give
their best performance at 10$^{\circ}$C ambient temperature. In future, a similar work will 
be carried out for the development of CZT-Imager payload, ASTROSAT.

On 30th January, 2009, CORONAS-PHOTON has been successfully launched and all the RT-2 
instruments, including RT-2/CZT-CMOS, are working with the expected performance. Details of the 
on-board calibration and data analysis will be published elsewhere.

\begin{acknowledgements}

TBK thanks RT-2/SRF fellowship (ISRO) which supported his research work. The authors are 
thankful to scientists, engineers and technical staffs from TIFR/ ICSP/ VSSC/ ISRO-HQ for 
various supports during RT-2 related experiments.

\end{acknowledgements}

{}
\end{document}